\documentclass[conference]{IEEEtran}
\usepackage{lineno}
\usepackage[hidelinks]{hyperref}

\usepackage{graphicx}
\usepackage{siunitx}
\DeclareGraphicsExtensions{.png,.eps,.ps,.pdf,.jpg}
\usepackage{url}
\usepackage[keeplastbox]{flushend}
\usepackage{color}
\usepackage{epstopdf}
\usepackage{float}
\usepackage{tikz}
\usepackage{pgf-pie}
\usepackage{diagbox, eqparbox, hhline}
\usepackage{pgfplots}
\usepackage{xcolor}
\usepackage{colortbl}
\usepackage{amsmath}
\usepackage{cases}
\usepackage[normalem]{ulem}
\usepackage[pdf]{pstricks}
\usepackage{acronym}
\usepackage{subfigure}

\newcommand\mypar[1]{\noindent \textbf{#1}}

\definecolor{ownApplegreen}{rgb}{0.55, 0.71, 0.0}
\newcommand{\eq}[1]{Eq.~(\ref{#1})}

\acrodef{SDN}[SDN]{Software Defined Networking}

\begin{document}

\title{Energy Sustainable Paradigms and Methods \\ for Future Mobile Networks: a Survey}

\author{\IEEEauthorblockN{Nicola Piovesan\IEEEauthorrefmark{1}, Angel Fernandez Gambin\IEEEauthorrefmark{2}, Marco Miozzo\IEEEauthorrefmark{1}, Michele Rossi\IEEEauthorrefmark{2}, Paolo Dini\IEEEauthorrefmark{1}}
    \IEEEauthorblockA{\IEEEauthorrefmark{1}CTTC/CERCA, Av. Carl Friedrich Gauss, 7, 08860, Castelldefels, Barcelona, Spain
    \\\{npiovesan, mmiozzo, pdini\}@cttc.es}
    \IEEEauthorblockA{\IEEEauthorrefmark{2}DEI, University of Padova, Via G. Gradenigo, 6/B, 35131 Padova, Italy
    \\\{afgambin, rossi\}@dei.unipd.it}}

\maketitle

\begin{abstract}
In this survey, we discuss the role of energy in the design of future mobile networks and, in particular, we advocate and elaborate on the use of energy harvesting (EH) hardware as a means to decrease the environmental footprint of 5G technology. To take full advantage of the harvested (renewable) energy, while still meeting the quality of service required by dense 5G deployments, suitable management techniques are here reviewed, highlighting the open issues that are still to be solved to provide \mbox{eco-friendly} and \mbox{cost-effective} mobile architectures. Several solutions have recently been proposed to tackle capacity, coverage and efficiency problems, including: C-RAN, \ac{SDN} and fog computing, among others. However, these are not explicitly tailored to increase the energy efficiency of networks featuring renewable energy sources, and  have the following limitations: (i) their energy savings are in many cases still insufficient and (ii) they do not consider network elements possessing energy harvesting capabilities. In this paper, we systematically review existing {\it energy sustainable paradigms and methods} to address points (i) and (ii), discussing how these can be exploited to obtain highly efficient, energy \mbox{self-sufficient} and high capacity networks.
Several open issues have emerged from our review, ranging from the need for accurate energy, transmission and consumption models, to the lack of accurate data traffic profiles, to the use of power transfer, energy cooperation and energy trading techniques. These challenges are here discussed along with some research directions to follow for achieving sustainable 5G systems.
\end{abstract}


%%%%%%%%%%%%%%%%%%%%%%%%%%%%%%%%%%%%%%%%%%%%%%%%%%%%
% INTRODUCTION
%%%%%%%%%%%%%%%%%%%%%%%%%%%%%%%%%%%%%%%%%%%%%%%%%%%%

\section{Introduction}
We live in the digital era. Dematerialization is becoming a reality, humans and machines alike are globally connected through the Internet. ITU estimated that $750$ million households are online and that there exist almost as many mobile subscribers as people in the world (around $6.8$ billions)~\cite{itu2016ict}. The trend is of a further increase in the traffic demand, in the number of offered and connected devices, especially mobile. The forecast in~\cite{cisco2017} is of an annual traffic growth rate of $53\%$, for the mobile traffic alone. This new era is undoubtedly opening up new possibilities for individuals as well as new opportunities for businesses and organizations. However, the massive use of ICT is also increasing the level of energy consumed by the telecommunication infrastructure and its footprint on the environment. In a report of 2013, the Digital Power Group~\cite{mills2013cloud} has calculated that $10\%$ of the worldwide electricity generation is due to the ICT industry, which surpasses of more than $50\%$ that of the avionic one. The report also highlights that the ICT energy consumption Compound Annual Growth Rate is of around $10\%$. In fact, forecasts for 2030 are that $51\%$ of the electricity consumption and $23\%$ of the carbon footprint by human activity will be due to ICT~\cite{andrae2015global}. Hence, any future development in the ICT technology and in its infrastructure has definitely to cope with their environmental sustainability.

Besides such increment in the demand, the ICT industry has to solve an economical problem, since operators' Average Revenue Per Unit (ARPU) is decreasing every year. The case of Vodafone Germany is particularly striking: its ARPU has been shrinking annually by $6\%$ on average in the period \mbox{2000-2009} \cite{Fehske2011}. One of the reasons of this is the annual increase of the OPerational EXpenditure (OPEX) of its network. Energy has been dominating these costs: it has been calculated that the energy bill equals the cost of the personnel required to run and maintain the network, for a western Europe company in 2007~\cite{Fehske2011}. Considering the rise in the energy price during the last few years, we conclude that energy saving is key for the economical sustainability of ICT.

In this survey, we discuss the crucial role of energy in the design of future networks, paying special attention to {\it mobile networks}, which are growing the most, among all ICT sectors, in terms of number of subscribers, traffic demand, connected devices and offered services~\cite{cisco2017}. 
Several survey papers, e.g.,~\cite{Buzzi2016, Mahapatra2016}, have recently appeared on these subjects, offering a thorough review of existing techniques and open issues. Nevertheless, existing solutions still have the following limitations: their energy savings are still insufficient and most of the research is still in a preliminary stage, they do not discuss the integration of energy harvesting capabilities into future networks and, in turn, energy \mbox{self-sustainability} is marginally addressed. The aim of the present survey is to fill these gaps, especially focusing and elaborating on the use of energy harvesting technology (including renewables) as a means to decrease the environmental footprint and OPEX of future mobile networks. We advocate that environmental energy can be scavenged through dedicated harvesting hardware, so as to power mobile system elements like base stations, mobile terminals and sensors. New network design paradigms, along with scenarios for future mobile networks and suitable network management techniques are also introduced, by reviewing the present literature and highlighting the open issues that are still to be solved to provide \mbox{eco-friendly} and \mbox{cost-effective} mobile networks. So, this survey approaches the existing literature from a different angle, aiming at energy harvesting and \mbox{self-sufficient} systems and, based on this take, trying to make some order in the different algorithms and concepts that were proposed so far, while identifying missing functionalities and discussing possible ways forward.

The paper is organized as follows. In Section~\ref{sec:big_picture}, we introduce the envisaged scenarios for future sustainable mobile networks, and elucidate the main objectives of this survey. The different sources of energy consumption are classified in Section~\ref{sec:energy_consumption}. Energy efficiency techniques, for both end-devices and network nodes, are reviewed in Section~\ref{sec:energy_efficiency}. In Section~\ref{sec:WPT}, we analyze the possibility of wirelessly transferring energy to \mbox{end-devices}. New network design paradigms, called energy cooperation and energy trading, are respectively described in Sections~\ref{sec:EnCoop} and~\ref{sec:EnTrad}. There, it is shown that network nodes can collaborate for energy \mbox{self-sustainability} and even trade some energy with the electrical grid to make profit. Our final remarks are given in Section~\ref{sec:conclusions}.

%%%%%%%%%%%%%%%%%%%%%%%%%%%%%%%%%%%%%%%%%%%%%%%%%%%%
% THE BIG PICTURE
%%%%%%%%%%%%%%%%%%%%%%%%%%%%%%%%%%%%%%%%%%%%%%%%%%%%

\section{The big picture}
\label{sec:big_picture}

In this section, we present our reference scenario and design principles for future sustainable mobile communications systems, often referred to as 5G, which constitute the basis for our discussion in the rest of the paper.

Current trends anticipate that 5G mobile networks will be composed of {\it ultra dense} deployments of heterogeneous Base Stations (BSs)~\cite{lopez2015}, where BSs using different transmission powers coexist to provide the 1000x network capacity increase that is required by 2020. Accordingly, the traditional macro cell layer will be complemented or replaced with multiple overlapping tiers of smaller cells, which extend the system capacity, thanks to a higher spatial reuse and to a better spectral efficiency.
Despite such benefits, researchers have already identified new issues raised by an ultra dense scenario, such as: user association and mobility management, interference management and mitigation, macro cell offloading, and energy saving~\cite{kamel2016}. Also, 5G subscribers will be equipped with a large and diverse set of devices and BSs may need to support high-rate mobile equipment (such as smartphones and laptops)~\cite{Andrews2014} as well as a huge number of \mbox{low-rate} devices (such as environmental or wearable sensors)~\cite{gubbi2013}, as envisaged by the Internet of Things (IoT) paradigm. 
This makes new generation networks challenging to operate, control and monitor. Moreover, such systems are also very demanding in terms of energy consumption from the power grid, due to their high capacity requirements. 
Different architectural designs have been proposed for next generation mobile networks including: 1) \mbox{Cloud-RAN} (C-RAN)~\cite{peng2016recent} and ~\cite{agiwal2016next}, 2) Software Defined Networks (SDN)~\cite{fundation2012}, 3) Network Function Virtualization (NFV)~\cite{li2015} and 4) Fog Computing~\cite{peng2015fog} and~\cite{hung2015architecture}. All these proposals rely on the cloud principle of sharing storage and computing resources. Moreover, they enable control and data plane decoupling and entail a pure software implementation of network functions, which may be opportunistically placed in different network elements.  

Such architectural proposals offer higher flexibility and scalability to operate, control and monitor new generation networks, however, a big effort is still necessary to reduce their energy demand. In fact, they are not specifically designed to reduce the energy consumption of 5G networks and to potentially make them energy \mbox{self-sufficient}.
Based on the literature reviewed in this survey, we acknowledge that there are studies that try to improve the energy efficiency of these designs. The surveys~\cite{Buzzi2016,Mahapatra2016} offer a comprehensive review of these methods together, including as list of open issues to be addressed. Nevertheless, we have identified the following limitations of existing schemes: (i) their energy savings are still insufficient and most of the research is still in a preliminary stage, (ii) they do not involve energy harvesting capabilities and energy self-sustainability. In this paper, we go beyond energy efficiency and provide a thorough review of existing {\it energy sustainable paradigms and methods} to address points (i) and (ii) to obtain (nearly) energy \mbox{self-sufficient} and high capacity networks. Towards this end, we advocate gathering environmental energy through dedicated harvesting hardware to supply 5G system elements (BSs, mobile devices, sensors, etc.). This translates into OPEX savings and into a reduction of the environmental footprint of ICT. The CAPital EXpense (CAPEX) can also be reduced~\cite{Piro2013} through the adoption of BSs with a small {form-factor}, as these require smaller energy amounts to be operated and this lessens the requirements in terms of harvesting and energy storage capabilities.

\begin{figure*}[]
	\centering
	\includegraphics[scale = 0.6]{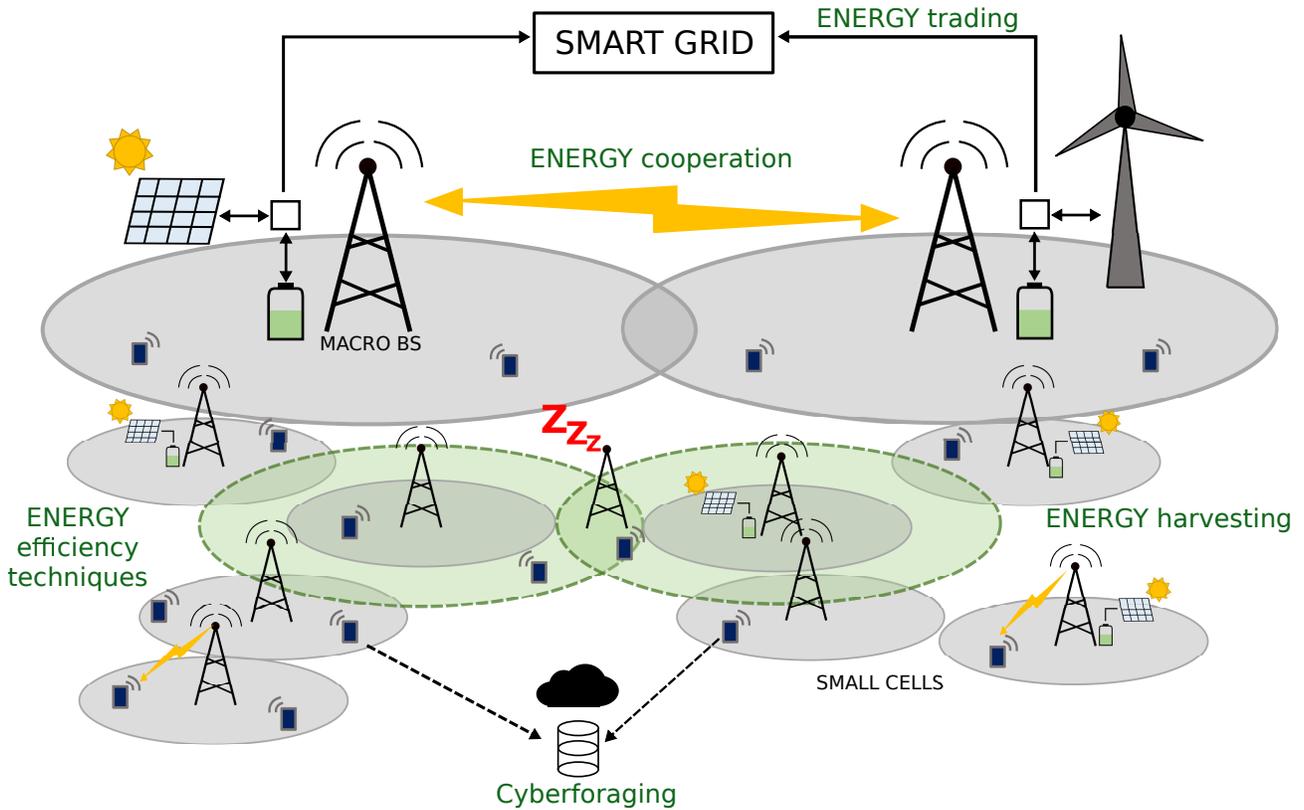}
	\caption{Diagram illustrating the main elements of future sustainable 5G networks. Small cells are utilized to increase the system capacity, energy harvester and energy storage devices ensure energy sustainability, while a range of technologies including cell zooming, cloud and fog computing help reduce the energy consumption of network elements. Energy cooperation, trading and transfer are accounted to balance the energy reserve across base stations and devices.}
	\label{fig:BP}
\end{figure*}

Fig.~\ref{fig:BP} illustrates our reference scenario, which includes base stations, mobile devices, sensors, energy harvesters and energy storage devices. Small cells are utilized to increase the system capacity, energy harvesters and energy storage devices ensure energy sustainability, while a range of technologies, methods and procedures including cell zooming, sleep modes and cyberforaging help reduce the energy consumption of network elements (energy efficient techniques in the figure). Energy cooperation, trading and transfer are utilized to balance the energy reserve across base stations and devices.

{\it Energy harvesting technology} will entail a higher management complexity. In fact, environmental energy, such as solar and wind, is inherently erratic and intermittent, which may cause a fluctuating energy inflow and produce service outage. A proper control of how the energy is drained and balanced across network elements is therefore necessary for a \mbox{self-sustainable} network design. 
The flexibility introduced by the cloud principles into the 5G architecture will definitely support the design of optimal strategies for network energy management.
Moreover, sustainable design of 5G systems shall rely on a set of procedures enabling energy efficient communication, such as cell sleeping and zooming, \mbox{Device-to-Device} (D2D) communication, cyberforaging and \mbox{energy-aware} communication hardware design. System \mbox{self-sustainability} may also benefit from other techniques such as wireless energy transfer (to prolong the battery life of \mbox{end-user} devices), energy cooperation (to efficiently exchange the harvested energy among BSs) and even energy trading (to sell/buy some of the energy to/from the power grid). These methods are detailed in Sections~\ref{sec:energy_efficiency}, \ref{sec:WPT}, \ref{sec:EnCoop} and~\ref{sec:EnTrad}. In each of these sections, we first review the most relevant techniques from the literature to then discuss open issues and provide suggestions for future research.  

At last, we remark that {\it energy storage devices} will play a key role in the design of sustainable mobile networks. In the rest of this survey, and mainly in Sections~\ref{sec:WPT}, \ref{sec:EnCoop} and~\ref{sec:EnTrad}, different uses of the accumulated energy are discussed. As identified in~\cite{Bayram2014}, the energy gathered from ambient sources can be utilized for: (i) improving power grid optimization in bulk power production, (ii) balancing the power system operations in the presence of intermittent renewable generation, (iii) helping to defer \mbox{capital-intensive} upgrades in the transmission and distribution grids and (iv) providing ancillary services to power grid operations. These points are much better addressed and optimized when energy storage is deployed. 

%%%%%%%%%%%%%%%%%%%%%%%%%%%%%%%%%%%%%%%%%%%%%%%%%%%%
% ENERGY CONSUMPTION
%%%%%%%%%%%%%%%%%%%%%%%%%%%%%%%%%%%%%%%%%%%%%%%%%%%%

\section{Energy consumption of network elements}
\label{sec:energy_consumption}

Before delving into the description of the techniques to make the network energy efficient and self-sufficient, next we review the main achievements in power consumption measurement and models for base stations and \mbox{end-devices}. 

\subsection{Base stations}
5G base stations can be classified into two main groups, depending on transmission power and coverage range.

\begin{itemize}
	\item[1)] \textbf{Macro BS}: with transmission power of about $\SI{40}{\watt}$ for devices with bandwidth of $\SI{20}{\mega\hertz}$ and $\SI{80}{\watt}$ for LTE-A devices with $\SI{40}{\mega\hertz}$~\cite{3gpp-36942}. Their communication range reaches up to a few kilometers and they are usually installed in building rooftops.
	\item[2)] \textbf{Small BS}: with transmission power ranging between $\SI{0.05}{\watt}$ and $\SI{6}{\watt}$. They can be further classified into micro, pico and femto BSs. Micro and pico BSs cover small to medium areas with dense traffic ({\it hotspots}) such as  shopping malls, residential areas, hotels, or train stations. The typical range of a micro/pico BS spans from a few hundred meters up to one kilometer. Femto cells are designed to serve smaller areas such as private homes or indoor spaces. The range of femto cells is typically only a few meters and they are generally wired to a private cable broadband connection or to a home digital subscriber line~\cite{Hasan2011green}. Small cells can be installed in street furniture like lampposts or traffic lights due to their small form factor. 
\end{itemize}

The power consumption at full system load of the different types of BSs can range from about $\SI{6}{\watt}$ for a femto BS to $\SI{1}{\kilo\watt}$ for a macro BS~\cite{Auer2013earth,deruyck2012,Sabella2016}. Typically, this power consumption is modeled as the sum of a static value and a dynamic and \mbox{load-dependent} value~\cite{Auer2011,Lorincz2012}:
\begin{equation}
\label{eq:BS_power}
P_{\rm BS} = \begin{cases}N_{\rm TRX}\cdot (P_0 + \alpha P_{\rm out}) , & 0 <P_{\rm out}\leq P_{\max} \\ N_{\rm TRX} \cdot P_{\rm sleep}, & P_{\rm out}=0 \end{cases}
\end{equation}
where $N_{\rm TRX}$ is the number of transmit/receive chains, $P_0$ is the BS power consumption at zero Radio Frequency (RF) output power, $\alpha$ is the slope of the load dependent power consumption curve, $P_{\rm out}$ is the \mbox{load-dependent} part of the RF output power and $P_{\max}$ is the value of $P_{\rm out}$ at maximum load.

Table~\ref{tbl:BS_param} specifies the load dependencies of the different BS types~\cite{Auer2013earth}. The power consumed by a macro BS increases much more with the traffic load than that of a small BS. This is due to the high consuming power amplifier that macro BSs use to cover wide areas, whereas small cells need amplifier designs for much lower coverage and, consequently, lower energy consumption figures.
Remarkably, $P_0$ represents a significant part of the total energy consumed by any BS and, due to this, researchers have investigated the use of sleep modes during low traffic periods. 
Moreover, it is expected that $P_0$ and $P_{\rm sleep}$ of new sites will be reduced by about $8\%$ on average thanks to recent technological advances~\cite{Sabella2016}, thus further decreasing the BS energy cost during low traffic periods.

\begin{table}[h!]
\centering
\caption{Power model parameters (from \cite{Auer2013earth})}
\label{tbl:BS_param}
\begin{tabular}{l | c c c c c}
\hline
\textbf{BS type}	& $N_{\rm TRX}$ & $P_{\max} [\SI{}{\watt}]$	& $P_0 [\SI{}{\watt}]$ & $\alpha$ & $P_{\rm sleep} [\SI{}{\watt}]$ \\   \hline
Macro	& 6			& 40.0		& 130.0 		& 4.7 	& 75.0 \\      
RRH		& 6			& 20.0		& 84.0 		& 2.8		& 56.0  \\
Micro	& 2			& 6.3  		& 56.0  		& 2.6		& 39.0\\   
Pico 		& 2			& 0.13 		& 6.8 		& 4.0 	& 4.3 \\ 
Femto	& 2 			& 0.05 		& 4.8 		& 8.0 	& 2.9 \\ \hline
\end{tabular}
\end{table}

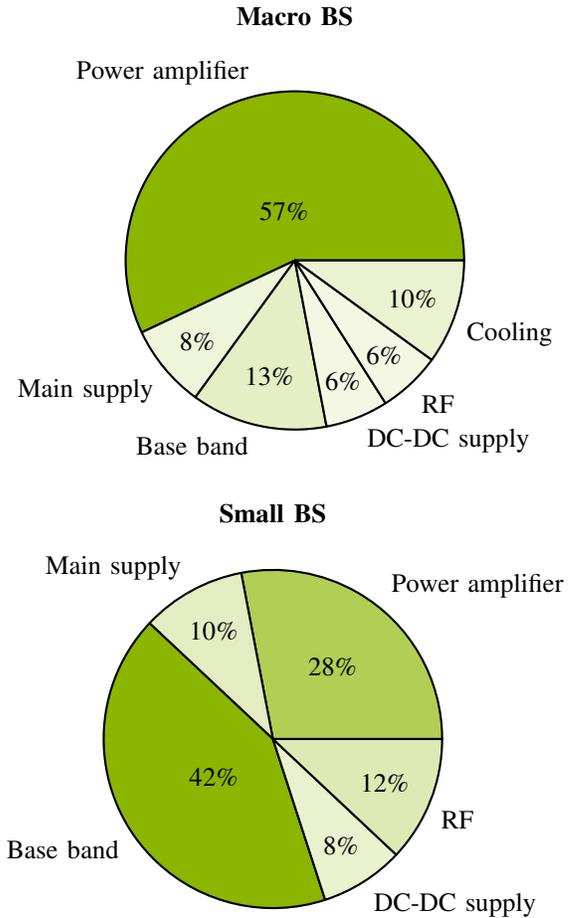
\begin{figure}[t]
 	\centering	
 	\begin{tikzpicture}[scale = 0.75]			
 	\pie[color = {ownApplegreen!100, ownApplegreen!14, ownApplegreen!23, ownApplegreen!11, ownApplegreen!11, ownApplegreen!18}]{57/Power amplifier, 8/Main supply, 13/Base band, 6/DC-DC supply, 6/RF, 10/Cooling}	
 	\node[above = 30mm, font=\bfseries]{Macro BS};	
 	\end{tikzpicture}
 	
 	\hspace{20mm}
 
 	\begin{tikzpicture}[scale = 0.75]			
 	\pie[color= {ownApplegreen!67, ownApplegreen!24, ownApplegreen!100, ownApplegreen!19, ownApplegreen!29}]{28/Power amplifier, 10/Main supply, 42/Base band, 8/DC-DC supply, 12/RF}	
 	\node[above = 27.5mm, font=\bfseries]{Small BS};	
 	\end{tikzpicture}
 	\caption{Comparison of base station energy consumption figures~\cite{Auer2013earth}.}
 	\label{fig:cons_comparative}
 \end{figure}

In Fig.~\ref{fig:cons_comparative}, we compare the energy drained by the various parts of macro and small BSs. According to~\cite{Auer2013earth}, the power amplifier of a macro BS dominates the total power consumption. For small BSs, the baseband processor has a higher impact. Gathering the baseband units of different BSs in a centralized pool, as done in \mbox{C-RAN} systems, may reduce the network energy consumption.

\subsection{End devices}

5G end-device are classified into two main categories: (i) {\it mobile phones} for \mbox{human-type} communications and (ii) {\it sensors} for \mbox{machine-type} communications. 

Smartphones integrate functionalities such as voice communications, Short Message Service (SMS), emailing, Web browsing and audio/video streaming and playback. These functionalities have a big impact on the battery life. The main hardware components of a smartphone are: the display (i.e., LCD panel, backlight, touchscreen and graphics subsystem), the radio module, the CPU, the RAM, the flash memory, the GPS and the audio module.
The impact of each of these components on the overall energy consumption depends on the operating mode. When no application is active, we can distinguish between two possible states: \emph{suspended} and \emph{idle}. In the \emph{suspended} state the application processor is idle, the display is off and the communications processor remains active in background to receive calls or messages. In this case, the component draining the highest amount of energy is the radio module. The \emph{idle} state is similar to the suspended state but the display is on and the graphics subsystem becomes the most energy demanding element.
Instead, when an application actively uses the device, the amount of energy drained by the different components highly depends on the usage scenario. For example, during a phone call the most energy draining component is the radio module, whereas, if a video is being played out, it is the display that consumes the most. In Table~\ref{tbl:smartphones}, the average power consumption figures (excluding the backlight) for three different smartphones are shown~\cite{Carroll2010}. From these results we note that the minimum amount of power is used up when the smartphone is in the suspended state. Moreover, besides being application dependent, the total energy consumption also (and strongly) depends on the specific vendor. 
In Table~\ref{tbl:backlight}, the variation of the power consumption due to the brightness level is provided for one of the smartphones of Table~\ref{tbl:smartphones}. In particular, the backlight can increase the consumption from $\SI{15}{\milli\watt}$ to more than $\SI{1000}{\milli\watt}$ depending on the application in use. Further details can be found in~\cite{Carroll2010}.

\begin{table}[t]
\centering
\caption{Power consumption of smartphone hardware components, from~\cite{Carroll2010}.}
\label{tbl:smartphones}
\begin{tabular}{l | c c c}
\hline
 & \multicolumn{3}{c}{Power consumption (\SI{}{\milli\watt})} \\
\textbf{Type of activity} & Freerunner & G1 & N1\\   \hline
Suspended	& 103.2	& 26.6 	& 24.9\\    
Idle			& 333.7	& 161.2	& 333.9\\   
Phone call		& 1135.4 	& 822.4 	& 746.8\\    
Email (cell) 	& 690.7 	& 599.4 	& -\\    
Email (WiFi)	& 505.6	& 349.2 	& -\\    
Web (cell)		& 500.0	& 430.4 	& 538.0\\    
Web (WiFi)	& 430.4 	& 270.6 	& 412.2\\    
Network (cell)	& 929.7	& 1016.4	& 825.9\\   
Network (WiFi)	& 1053.7	& 1355.8 	& 884.1\\    
Video		& 558.8	& 568.3 	& 526.3\\    
Audio		& 419.0	& 459.7 	& 322.4\\    \hline
\end{tabular}
\end{table}

\begin{table}[t]
\centering
\caption{Power consumption figures for the N1 smartphone from~\cite{Carroll2010}.}
\label{tbl:backlight}
\begin{tabular}{ l | c c c}
\hline
\textbf{Type of activity}	&	Min power (mW)	& Max power (mW)\\   \hline
Idle			& 38.0	& 257.3 \\    
Phone call		& 16.7	& 112.9\\   
Web			& 164.2	& 1111.7\\     
Video		& 15.1 	& 102.0\\     \hline
\end{tabular}
\end{table}

A power consumption model for the radio module is proposed in~\cite{Jensen2012} for the first generation of LTE. It considers the receive/transmit power levels, the uplink/downlink data rate, and the Radio Resource Control (RRC) mode. The authors observe that the uplink transmit power and the downlink data rate are the factors that affect the most the overall power consumption, while the contributions due to the uplink data rate and to the downlink receive power are small.
A more detailed model is proposed in~\cite{lauridsen2013empirical}, where the impact of the cell bandwidth and of the Discontinuous Reception (DRX) mode, introduced with the second generation of LTE to enable radio sleep modes, are also considered. In this study, it is concluded that the power consumption of the LTE radio module increases of about three times when going from a bandwidth of $\SI{10}{\mega\hertz}$ to $\SI{15}{\mega\hertz}$. Furthermore, two sleep modes are available: a light sleep mode consuming 0.57 W and a deep sleep mode consuming 29 mW. The measurements provided by the authors show that sleep modes allow reducing the power consumption from $1/1.8$ (light sleep mode) to $1/35$ (deep sleep mode) with respect to that in the \mbox{idle-state} (radio active with no data reception).

The second class of terminals are the sensor nodes, which are usually composed of different elements such as the \mbox{micro-controller}, the sensor, the transceiver, the memory, the interface, the DC-DC converter and the battery~\cite{hesse2016towards}. 
Their power consumption is strictly related to the type of components and to the specific sensing application. Basically, a sensor executes two main tasks: collecting measurements from the environment and transmitting data. During each of these a number of components operates in active mode. The energy consumption of a task depends on the power consumed by the active components and on the duration of the task. For the sake of illustration, we consider the example in~\cite{hesse2016towards}, whose measurements are shown in Table~\ref{tbl:sensor}. The sensor node is equipped with a controller, temperature, humidity and barometric pressure sensors, an ambient light sensor and a Bluetooth radio transceiver. Other components such as the DC/DC converter or the battery are not shown in the table. This sensor node consumes from a minimum of $\SI{0.09}{\milli\ampere}$, when all the components are inactive, to a maximum of $\SI{14}{\milli\ampere}$, when all the components are active. The most energy hungry component is the radio transceiver.

\begin{table*}[t]
\centering
\caption{WSN node power consumption figures from~\cite{hesse2016towards}.}
\label{tbl:sensor}
\begin{tabular}{ l | l | l}
\hline
Component						& Active mode		 & Inactive mode\\   \hline
Controller (SAML21)	& \SI{0.3515}{\milli\ampere} & \SI{0.0060}{\milli\ampere} (standby) -- \SI{0.1968}{\milli\ampere}  (Idle) \\    
Temp./Hum./Press. sensor (BME280) & \SI{0.468}{\milli\ampere} & \SI{0.0005}{\milli\ampere} (standby)\\   
Ambient light sensor (BH1715)	& \SI{0.190}{\milli\ampere} & \SI{0.0010}{\milli\ampere} (powerdown)\\     	
Bluetooth TRX (BLE112) & \SI{13}{\milli\ampere} & \SI{0.0809}{\milli\ampere} \\     \hline
\end{tabular}
\end{table*}

\subsection{Main outcomes}

Base station power consumption models are based on the assumption that hardware and software have been jointly designed and implemented. However, this fact does not longer hold true for future mobile networks based on software virtualization. In that case, measurements indicating the CPU usage by different (virtualized) BS functions shall be performed considering the right hardware, which may differ for each function according to the actual execution point (e.g., in the cloud or at the network edge).

Power consumption models for smartphones shall as well be improved. Current models are mainly based on old (2G and 3G) mobile networks. A new analysis and experimental measurements would be needed to evaluate the evolution of the hardware components as well as of the new software features offered by new generation smartphones. The availability of the circuit schematics is key to define a power consumption model for the hardware components, but this information is seldom provided by the phone vendors (a notable exception is offered by OpenMoko~\cite{openmoko2017}). 

From our literature analysis, we see that different brands have totally different power requirements. Hence, mobile phones from different manufacturers shall be examined and ranked according to their energy consumption figures. Also, the energy expenditure of smartphones strongly depends on the owner's usage pattern, which shall be determined to define accurate energy consumption models. 

Few works in the literature study the power consumption of sensor devices. New measurements would be needed to develop accurate sensor models. Furthermore, those models should include a detailed analysis of the microcontroller consumption, considering important features like interrupt handling and DMA operations.

As a final remark, we have also identified a lack of research papers on battery and energy storage models for \mbox{end-devices} and BSs, which are key to a proper design of future sustainable mobile networks.

%%%%%%%%%%%%%%%%%%%%%%%%%%%%%%%%%%%%%%%%%%%%%%%%%%%%
% ENERGY EFFICIENCY TECHNIQUES
%%%%%%%%%%%%%%%%%%%%%%%%%%%%%%%%%%%%%%%%%%%%%%%%%%%%

\section{Energy efficiency techniques}
\label{sec:energy_efficiency}

Next, we concentrate on the techniques to reduce the energy consumption of the mobile system as a whole. Energy Efficiency (EE) is the fundamental brick of any sustainable design and defines the key methods that are to be either enhanced or brought forward when integrating energy harvesting sources. We refer to EE as a set of functions/methods conceived to reduce the energy requirement for a given level of service. EE can be quantified by the ratio between the amount of data successfully delivered (in $\SI{}{bit/s}$) and the total energy spent in such transmission (in $\SI{}{\watt\hour}$ or $\SI{}{\joule}$).  

Several surveys have been written to discuss on the energy efficiency of the mobile system. Sources of inefficiencies in the network are described in \cite{Davaslioglu2014}, where some potential improvements are also suggested. The authors of \cite{Mahapatra2016} provide an extensive description of \mbox{energy-aware} mechanisms at each protocol layer of the communication stack, including energy efficient hardware design principles. In this section, instead, we only concentrate on the energy efficient techniques at the network and \mbox{end-user} side, which can enable an intelligent use of the harvested ambient energy and support the system \mbox{self-sustainability}.  

\subsection{Network energy efficiency}

The EE techniques that are exploited to decrease the energy footprint of BSs fall under two categories: 1) {\it sleep modes}, to selectively switch off some of the radio units (according to the traffic profile) and 2) {\it cell zooming}, to adapt the coverage range of BSs to cover areas where BSs are asleep and perform load balancing. These techniques are analyzed in the following.\\

\mypar{1) Sleep modes}: cellular networks are dimensioned to support traffic peaks, i.e., the number of BSs deployed in a given area should be able to provide the required Quality of Service (QoS) to the mobile subscribers during the highest load conditions. However, during \mbox{off-peak} periods the network may be underutilized, which leads to an inefficient use of spectrum resources and to an excessive energy consumption (note that the energy drained during low traffic periods is non-negligible due to the high values of $P_0$ in \eq{eq:BS_power}). For these reasons, sleep modes have been proposed to dynamically turn off some of the BSs when the traffic load is low. This has been extensively studied in the literature, considering different problem formulations \cite{Han2016}. As BSs cannot serve any traffic when asleep, it is important to properly tune the enter/exit time of sleep modes to avoid service outage. Moreover, when a BS is switched on/off, there is an incurred energy cost that should not be ignored. This is tackled in~\cite{yu2016minimizing} by considering BSs state transitions over time in the optimization problem, such that the overall BSs switching energy cost is minimized.

The authors of~\cite{Zhang2013energy} propose centralized and distributed clustering algorithms to cluster those BSs exhibiting similar traffic profiles over time. Upon forming the clusters, an optimization problem is formulated to minimize their power consumption. Optimal strategies are found by brute force, since the solution space is rather small and its complete exploration is still doable. 
A similar approach is presented in~\cite{samarakoon2016dynamic} where a dynamic switching on/off mechanism locally groups BSs into clusters based on location and traffic load. The optimization problem is formulated as a \mbox{non-cooperative} game aiming at minimizing the BS energy consumption and the time required to serve their traffic load. Simulation results show energy costs and load reductions while also provide insights of when and how the cluster-based coordination is beneficial.

User QoS is added to the optimization problem in~\cite{Cai2013cross}. In this case, as the problem to solve is \mbox{NP-hard}, the authors propose a suboptimal, iterative and \mbox{low-complexity} solution. The same approach is used in~\cite{Zhu2014qos, Tao2016energy, Bousia2012energy, ghazzai2017green}, playing with the \mbox{trade-off} between energy consumption and QoS. 
The Quality of Experience (QoE) is included in~\cite{Yuan2015qoe}, where a dynamic programming switching algorithm is put forward. The user QoE is utilized in place of standard network measures such as delay and throughput. Other parameters that have been considered are the channel outage probability (also referred to as coverage probability), i.e., the probability of guaranteeing the service to the users located in the worst positions (e.g., at the cell edge) and the BS state stability parameter, i.e., the number of on/sleep state transitions. For instance, a set of BS switching patterns engineered to provide full network coverage at all times, while avoiding channel outage, is presented in~\cite{Han2013energy}. The coverage probability, along with power consumption and energy efficiency metrics, are derived using stochastic geometry in~\cite{Wang2015energy, He2014stochastic, soh2013energy}. A similar approach is considered in~\cite{Jia2015resource}, where \mbox{closed-form} expressions of coverage probability and average user load are attained through stochastic geometry. Optimal resource allocation schemes are proposed to minimize power consumption and maximize coverage probability in a Heterogeneous Network (HetNet), and are validated numerically. According to the BS state stability concept, a \mbox{bi-objective} optimization problem is formulated in~\cite{Liu2015base} and solved with two algorithms: (i) near optimal but not scalable, and (i) with low complexity, based on particle swarm optimization.
The QoE is also affected by the UE positions according to the channel propagation phenomena. To this respect, in~\cite{Bousia2012ICC} the selection of the BSs to be switched off is taken in order to provoke less impact to the UEs' QoE according to their distance to the handed off BSs.

In order to support sleep modes, neighboring cells must be capable of serving the traffic in off areas. To achieve this, proper \textit{user association} strategies are required. In a scenario where sleeping techniques are not applied, each user is associated with the BS that provides the best Signal to Interference plus Noise Ratio (SINR). However, when BSs can go to sleep, user association is more complex and requires traffic prediction as well as very fast \mbox{decision-making}. Otherwise, users may suffer a deterioration of their QoS. A framework to characterize the performance (outage probability and spectral efficiency) of cellular systems with sleeping techniques and user association rules is proposed in~\cite{Tabassum2014downlink}. In this paper, the authors devise a user association scheme where a user selects its serving BS considering the maximum expected channel access probability. This strategy is compared against the traditional maximum \mbox{SINR-based} user association approach and is found superior in terms of spectral efficiency when the traffic load is inhomogeneous. User association mechanisms that maximize energy efficiency in the presence of sleep modes are addressed in~\cite{Zhu2014energy}. There, a downlink HetNet scenario is considered, where the energy efficiency is defined as the ratio between the network throughput and the total energy consumption. Since this leads to a highly complex integer optimization problem, the authors propose a Quantum particle swarm optimization algorithm to obtain a suboptimal solution. Moreover, a problem that jointly considers energy cost and \mbox{flow-level} performance, such as file transfer delay, is formulated in~\cite{Son2011base}. This formulation is decomposed  into two subproblems: user association and BS operation. For the user association, an optimal policy is derived, also devising a distributed implementation. For the BS operation, some \mbox{low-complexity} algorithms are proposed. 

Mobile network operators (MNOs) cooperation is exploited in~\cite{oikonomakou2017evaluating} where a switching off strategy is implemented through a roaming cost based on user association to offload traffic and eventually defines the operational state of the BSs.
Similarly, in~\cite{Bousia2016Sharing} the switch on/off problem for clusters of BSs has been modeled with a non-cooperative game with complete information algorithm. The game is played by the MNOs for estimating the switching-off probabilities that reduce their expected financial cost when roaming the traffic. The proposed scheme improves both the network energy efficiency and the cost. Results also provide understandings on the MNOs behavior as function of the roaming cost.
An auction-based switching off solution has been proposed in~\cite{Bousia2016Auction} where the macro BSs owned by different MNOs can offload traffic to third party small BSs. A \mbox{multi-objective} auction framework has been used to opportunistically utilize the small BSs. The proposed solution considers different bidding strategies representing different levels of tolerance respect to the QoE that the MNOs want to provide to their UEs. Simulation results show improvements for throughput, energy efficiency and cost savings, providing also guidelines concerning the behaviors that the MNOs should follow in the auction.
Finally, cooperation between MNOs in a C-RAN architecture is analyzed in \cite{Vincenzi2017cooperation}. The authors propose a novel scheme based on coalitional game theory to identify the potential room for cooperation among MNOs that provide service to the same area. Simulation results show that for the operators it is always more convenient collaborating, with profit gains ranging above 98\% when compared to the stand-alone case. 

\mypar{2) Cell zooming}: this method is also known as \textit{cell breathing}, it is complementary to the above user association techniques and has been introduced to fill the coverage gaps that may occur as BSs go to sleep. It amounts to adjusting the cell size according to traffic conditions, leading to several benefits: (i) load balancing is achieved by transferring traffic from highly to lightly congested BSs, (ii) energy saving through sleeping strategies, (iii) user battery life and throughput enhancements~\cite{Ismaiil2014reducing}. To compute the right cell size, cell zooming adaptively adjust the transmit powers, antenna tilt angles, or height of active BSs. There exists a large number of works that apply this approach to achieve energy savings in cellular networks. For instance, a cell zooming scheme, to be used in \mbox{two-tier} cellular networks with macro and small cells, is put forward in~\cite{Chung2015energy}. The considered formulation entails a Capacitated Facility Location Problem (CFLP), which is known to be \mbox{NP-hard}. Hence, the authors provide a practical implementation allowing BSs to be smartly switched on/off and filling coverage holes zooming in and out the active BSs. Further, centralized and distributed cell zooming algorithms are proposed in~\cite{Niu2010cell}, where a cell zooming server, which can be either implemented in a centralized or distributed fashion, controls the zooming procedure by setting its parameters based on traffic load distribution, user requirements, and Channel State Information (CSI). The same \mbox{server-based} solution can be found in~\cite{Ismaiil2014reducing}. A different approach is proposed in~\cite{Le2012qos}, where the authors design a BS switching mechanism based on a power control algorithm that is built upon \mbox{non-cooperative} game theory. A closed-form expression cell zooming factor is defined in~\cite{xu2017adaptive}, where an adaptive cell zooming scheme is devised to achieve the optimal user association. Then, a cell sleeping strategy is further applied to turn off light traffic load cells for energy saving. In general, most zooming scenarios entail a computationally intractable formulation, so affordable solutions based on iterative algorithms or heuristics abound in the literature, see, e.g.,~\cite{Hu2016green, Zhu2013qos}. 

Remarkably, cell zooming entails an increase in the transmit power of the active BSs, which leads to a higher energy expenditure for the BSs that are on. However, when used in combination with sleeping strategies, this leads to additional energy savings. Some researchers are oriented towards the study of sleeping schemes in conjunction with cooperative communication strategies for distributed antennas, also referred to as \textit{Coordinated Multi Point} (CoMP). This technique increases spectral efficiency and cell coverage without entailing a higher BS transmit power and reducing the \mbox{co-channel} interference. The authors of~\cite{Cili2012cell} prove the effectiveness of this approach in terms of energy and capacity efficiency when sleep modes are combined  with downlink CoMP. Despite these advantages, their results also reveal that imperfect downlink channel estimations and an incorrect CoMP setup can lead to energy inefficiency. A stochastic geometry analysis is presented in~\cite{He2014stochastic} to evaluate the energy efficiency performance of joint sleeping and CoMP in HetNets. The authors of this paper compare the coverage probability and the energy efficiency in scenarios with and without CoMP. Their results demonstrate that the combined use of CoMP and BS sleeping techniques can improve the energy efficiency and increase the coverage probability when compared with the sole use of sleep modes.

\begin{figure}[t]
	\centering
	\includegraphics[width=\columnwidth]{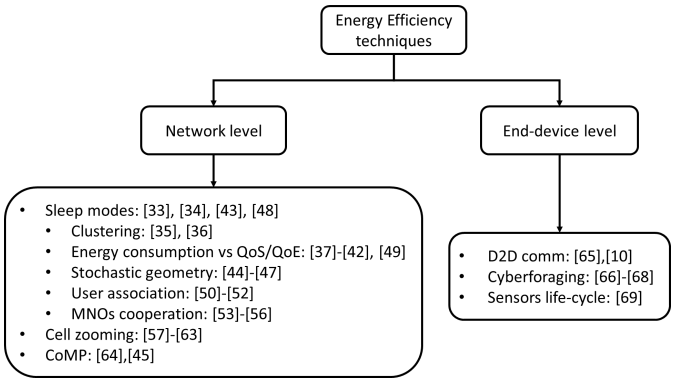}
	\caption{Energy efficiency techniques reviewed in this section.}
	\label{fig:ee_graph}
\end{figure}

\subsection{End-devices}

In the previous generation of cellular networks, mobile devices were used to make phone calls, the communicating peers were assumed to be far from one to another and mobile networks were designed with the premise of having full control at the infrastructure and mobile side. However, today's scenario has radically changed: data sharing between geographically close users does represent a reality, and devices are smarter and more powerful in terms of computation and memory. In this context, in 5G networks D2D communication can be exploited to increase network performance and decrease power consumption~\cite{Boccardi2014, Andrews2014}. The ability of two users to communicate directly reduces the number of wireless hops to one, thus increasing spectral efficiency (due to the reduced path loss), reducing latencies and limiting network signaling. Furthermore, this operation requires a smaller transmit power, since the receiving device is likely to be closer than the nearest BS. 

Another important point to consider is that mobile devices have limited battery duration and require to be recharged often. As discussed in Section~\ref{sec:big_picture}, one of the reasons for this is due to computational expensive tasks that are executed by mobile applications. Hence, enabling smartphones to offload their highly \mbox{energy-demanding} operations to nearby network servers~\cite{Barbarossa2014} is a viable solution to prolong their battery life. This strategy is known in the literature as \emph{computational offloading}~\cite{fernando2013mobile} or \emph{cyberforaging}~\cite{sharifi2012survey}. Cloud computing can be also exploited to support cyberforaging through different kinds of resources, such as infrastructures, platforms and software. The software executed by mobile devices is usually subdivided into modules. This makes it possible to only offload some specific modules to the remote servers. Cyberforaging can be static or dynamic. In the first case, the mobile device decides which modules to offload at the beginning of the software execution. In the second case, this decision is made at runtime according to the CPU load. The decision to offload a module is made taking into account the following aspects: (i) the number of operations to be executed, (ii) the time needed to offload, (iii) the required input. Moreover, the mobile device radio module influences offloading decisions by providing information on the power required to transmit the module to the external server and on the expected latency (i.e., the sum of the time taken to transmit the module to the server, the time the server spends executing the program, and the time required to send the results back to the mobile device). Network BS densification can potentially support cyberforaging by reducing the distance between mobile devices and small cells.

In the case of sensor nodes, the energy storage capacity is even more limited due to their very small form factor. The batteries represent the bottleneck for the lifetime of a wireless sensor network and for this reason energy efficient algorithms are key to prolong the system life-cycle. The literature on this topic is vast and has been mainly focused on the definition of energy efficient medium access control and routing protocols. For a detailed \mbox{state-of-the-art} on this topic, the reader is referred to~\cite{Rault2014}.

\subsection{Main outcomes}
\label{sec:ee_outcomes}

Figure~\ref{fig:ee_graph} summarizes the energy efficiency techniques reviewed in this section. The main findings are the following:
\begin{enumerate}
	\item Grouping BSs with similar traffic patters through clustering techniques provide valuable results when applied to BS sleep modes.
	\item Stochastic geometry has been vastly used to analyze the EE performance in switching on/off strategies.
	\item BS sleeping solutions shall be combined with other techniques such as user association, cell zooming and CoMP to ensure satisfactory network performance. 
	\item MNOs cooperation has been exploited through game theory and auction-based approaches with promising outcomes.
	\item More powerful devices allow D2D communications enhancing the network performance and saving energy. 
	\item Cyberforaging together with cloud computing prolong the end-user battery life through computational tasks offloading.
\end{enumerate}

Sleeping techniques have been widely investigated for cellular networks, but there are still some open problems to be solved. In the review of the literature we noticed that the traffic models are usually \mbox{over-simplified}, considering uniform traffic distributions and arrival patterns in all cells at all times.  
However, actual network traffic is dynamic and undergoes spatial and temporal fluctuations~\cite{Auer2013earth} due to the movement of UEs. Hence, accurate mobility models should be inferred from real data, and used to investigate the performance of sleep modes and cell zooming. 
Moreover, BS switching operations are usually modeled without considering activation frequency and time. Although the most recent BSs have been conceived for frequently entering sleep modes, most of the BSs that are still in use today were designed foreseeing only occasional switch on and off operations, as otherwise the failure rate of some of their parts would be too high~\cite{Wu2015energy}. Besides, fast switching operations can lead to a \textit{ping-pong} effect, which occurs when the service is handed over from one cell to another, but is quickly handed back to the original cell increasing control messages to the core network, leading to an increased energy consumption and to a decreased user QoS~\cite{Shin2015siesta}. This is more severe when there is a non-negligible BS activation time, as resources may be deactivated due to a temporary decrease in the load, and cannot be rapidly reactivated in response to a sudden increase of the same~\cite{elayoubi2011optimal}. These aspects are to be taken into account to avoid service outage in real world scenarios.

In the case of D2D communications, research challenges mainly concern the management of interference and the analysis of the involved parameters and tradeoffs. Transmission powers should be regulated to avoid interference with cellular communications (in case of inband D2D) or with other systems (in case of outband D2D). Furthermore, accurate channel information is essential to efficiently perform resource/power allocation and interference management. Since the introduction of D2D will increase the amount of channels that the devices need to estimate, the tradeoff between CSI accuracy and the resulting overhead should be analyzed. Other  open issues concern the design of energy efficient D2D protocols. As an example, the search for devices in close proximity is an energy expensive operation and the frequency at which this operation is performed highly affects the UE's power consumption. This involves an inherent tradeoff between detection performance and energy consumption.

Finally, the introduction of cyberforaging will lead to an increase in the uplink traffic. The corresponding downlink/uplink traffic statistics will differ from what we have today and will require a different partitioning (including dynamic allocation) of the capacity for the two channels.

%%%%%%%%%%%%%%%%%%%%%%%%%%%%%%%%%%%%%%%%%%%%%%%%%%%%
%  WIRELESS POWER TRANSFER
%%%%%%%%%%%%%%%%%%%%%%%%%%%%%%%%%%%%%%%%%%%%%%%%%%%%

\section{Wireless Power Transfer}
\label{sec:WPT}

\begin{figure}
 \centering
 \subfigure[WET]{\includegraphics[scale = 0.7]{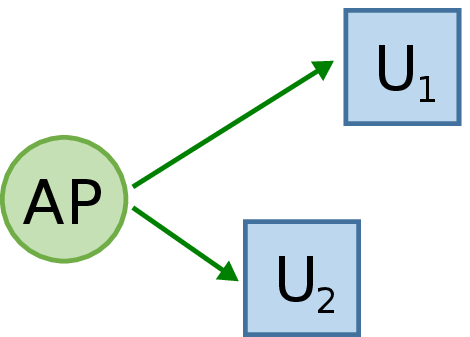}}
\subfigure[SWIPT]{\includegraphics[scale=0.7]{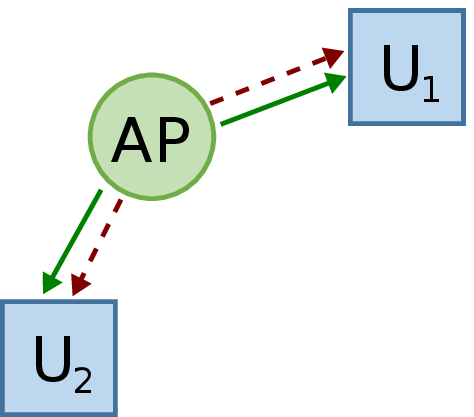}}
\subfigure[WPCN]{\includegraphics[scale=0.7]{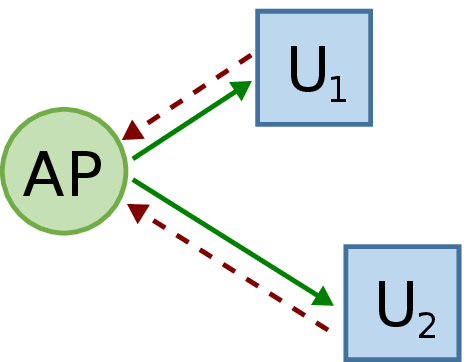}}
\caption{WPT architectures. Solid arrows represent energy transmission, whereas dashed arrows indicate information transmission.}  
\label{fig:wpt_arch}
 \end{figure}

In the following, we analyze wireless power transfer techniques. 
In the literature, this technology has been studied from different perspectives:
\begin{itemize}
\item[1)] \textbf{Wireless Energy Transfer (WET):} concentrating on the energy transfer from BSs to UEs (downlink);
\item[2)] \textbf{Simultaneous Wireless Information and Power Transfer (SWIPT):} where both energy and information are transferred in downlink;
\item[3)] \textbf{Wireless Powered Communication Network (WPCN):} where energy is transferred in downlink, while information is transferred in uplink.
\end{itemize}

In the following subsections, the main outcomes of these research efforts are discussed in some detail.

\subsection{Wireless Energy Transfer}
\label{subsec:WET}

First, we consider the case of a transmitter that wirelessly transfers energy to multiple receivers. In general, (power) senders and receivers are equipped with multiple antennas and the transmitted signal is modulated. The energy harvesting module at the receiving end is based on a rectifying circuit that is composed of a diode and a low pass filter. This circuit converts the received RF signal into a DC one.

According to~\cite{Bi2015}, the harvested energy per unit time is proportional to the received RF power. To improve it, one can increase the number of antennas at both transmitter and receiver, allowing a higher (combined) antenna gain. This solution, referred to as {\it energy beamforming}, effectively steers the transmit power towards a specific direction, with a subsequent improvement in the energy transfer efficiency. Furthermore, the width of the energy beam can be narrowed by increasing the number of antennas.

When considering the simultaneous charge of multiple energy receivers, a beamforming approach can lead to a \mbox{\textit{near-far}} problem, where the users close to the transmitter receive more energy than those located further away. Furthermore, the use of beamforming requires an accurate knowledge of the channel state at the transmitter, but in many cases energy transmitters are simple devices, which do not possess signal processing capabilities. Including such capabilities comes at the cost of an increase in the device energy consumption and in its processing time. The acquisition of the Channel State Information (CSI) is investigated in~\cite{Zeng2015}, where the channel reciprocity is exploited to design an efficient channel acquisition method for a \mbox{point-to-point} \mbox{Multi-Input} \mbox{Multi-Output} (MIMO) WET system. In this paper, the antenna weights are set through a training phase, which is formulated as an optimization problem for the case of uncorrelated fading channels. Optimal solutions  are derived for the special cases of MIMO Rayleigh and MISO Rician fading channels, with the aim of maximizing the net harvested energy at the energy receiver.

\subsection{Simultaneous Wireless Information and Power Transfer}

The SWIPT technique aims at transmitting energy and information through the same waveform, considering the fact that information signals also carry energy that can be harvested by an energy receiver.

Generally, information detection (ID) and energy harvesting (EH) receivers have different power sensitivities ($\SI{-10}{dBm}$ for EH, $\SI{-60}{dBm}$ for ID, according to~\cite{Bi2015}). This means that to work properly, EH receivers should be closer to the transmitter than ID receivers.

Since the design of the waveforms has a major impact on the performance of simultaneous energy and information transfer, a tradeoff between energy transmission and information transmission efficiencies has to be found. As for the energy transmission, the objective corresponds to maximizing the power transferred to the end user, whereas, for the information transmission, it is the transmission rate that has to be maximized. In the literature, this tradeoff is explored through the definition of \mbox{Rate-Energy (R-E)} regions, which contain all the feasible rate (\SI{}{bit/\second/\hertz}) and energy (\SI{}{\joule/\second}) pairs under a maximum transmit power budget. For any given technique, the optimal tradeoff between energy and information transfer rates is provided by the boundary of the corresponding \mbox{R-E} region, and depends on the receiver structure.
An ideal receiver, that jointly decodes information and harvests energy from the same signal, using the full signal power for both tasks, is physically infeasible. Thus, the following practical receiver designs are proposed~\cite{Bi2015}:

\begin{enumerate}
\item \textit{Time switching:} the transmitter sends data (ID) and energy (EH) using disjoint time slots. Within each time slot, the transmission can be optimized depending on its content (energy or information). The receiver, periodically switches between harvesting energy and decoding information.

\item \textit{Power splitting:} the transmitter sends a single waveform to carry energy and information. The receiver splits the received signal into two streams: one stream with power ratio $0 \leq \rho \leq 1$ is used for energy harvesting, the other, with power ratio $1-\rho$, is used to decode the data message.

\item \textit{Integrated receiver:} the received signal is at first converted into DC current and then split into two streams. This solution allows using a passive rectifier for \mbox{RF-to-baseband} conversion, which entails a lower energy usage when compared to the active mixer that is required by the information decoder of the previous technique.

\item \textit{Antenna switching:} this solution can be used when the receiver is equipped with multiple antennas. In this case, the receiver can use a number of antennas for energy harvesting and the remaining ones for information decoding. This simple solution reduces the hardware complexity at the receiver side, as it only needs to synchronize a switch.
\end{enumerate}

\begin{figure}
 \centering
 \subfigure[Time switching]{\includegraphics[scale = 0.5]{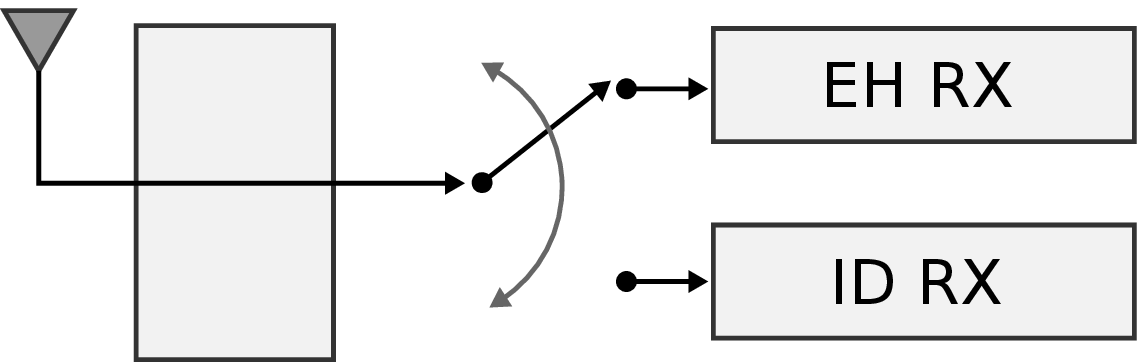}}
\subfigure[Power splitting]{\includegraphics[scale=0.5]{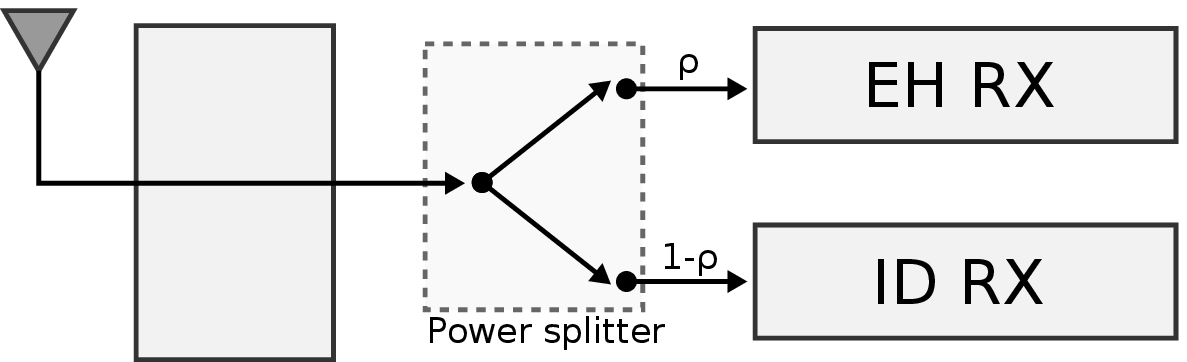}}
\subfigure[Integrated receiver]{\includegraphics[scale=0.5]{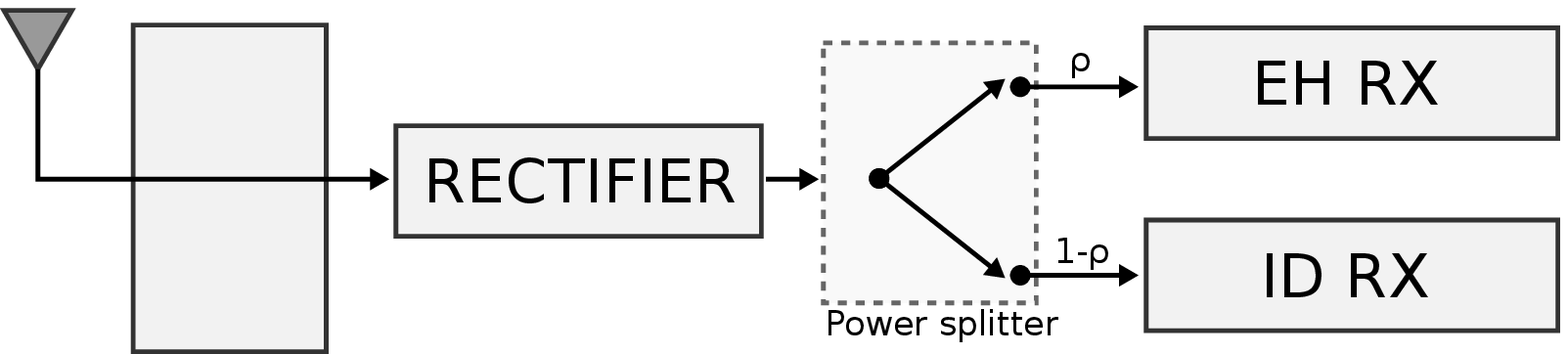}}
\subfigure[Antenna switching]{\includegraphics[scale=0.5]{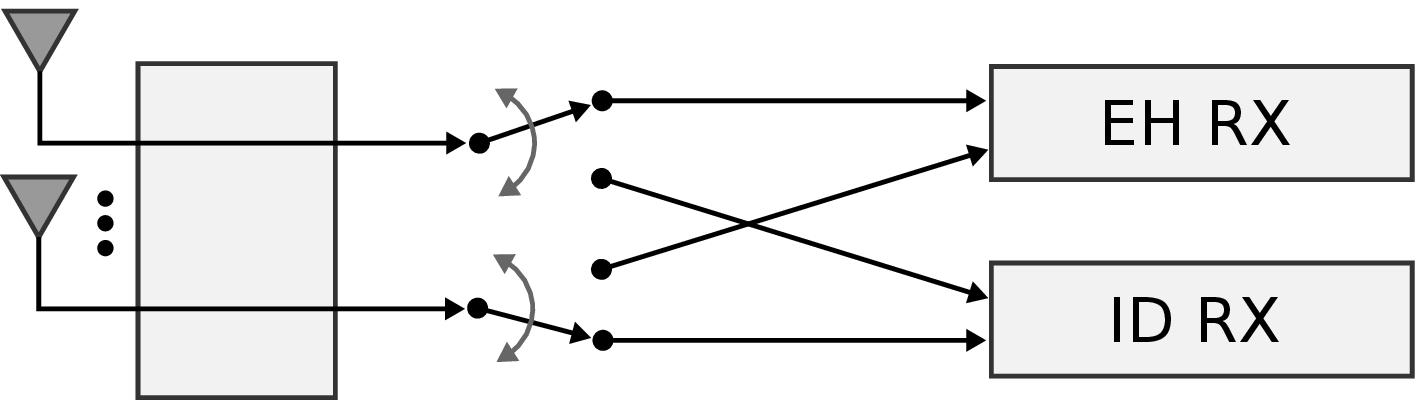}}
\caption{SWIPT receiver designs}  
\label{fig:wpt_swipt}
 \end{figure}

In~\cite{Zhang2013}, a MIMO wireless broadcast system is investigated. In the considered setup, there are three nodes, one transmitter, one energy harvesting receiver and another receiver that decodes information. The cases of (i) disjoint and (ii) \mbox{co-located} receivers are explored. In the first case, the two receivers see two different channels, while, in the second, they experience the same channel. For the MIMO link between the transmitter and the energy harvesting receiver, the amount of energy harvested is maximized through beamforming. For the MIMO link between the transmitter and the data decoder, the transmission rate is maximized through spatial multiplexing. The \mbox{R-E} region is computed to assess the optimal broadcasting policy in the case of simultaneous wireless power and information transfer. In scenario (i), where the receivers are disjoint, the beamforming strategy is demonstrated to be optimal when considering MISO links (between the transmitter and the EH/ID receivers). It is also shown that increasing the correlation between the two channels widens the \mbox{R-E} region, proving that an increase in the antenna correlation is beneficial. In scenario (ii), where the two receivers are \mbox{co-located}, the optimal strategy is spatial multiplexing.

An improvement is proposed by~\cite{Xiang2012}, where the robust beamforming problem in a MIMO SWIPT wireless broadcasting system is investigated under the assumption of imperfect channel state information at the transmitter. The objective is to maximize the \mbox{worst-case} harvested energy for the energy receiver, while guaranteeing that the information transmission rate is above a given threshold, for all the possible channel realizations. This amounts to a \mbox{non-convex} problem, which is relaxed into a \mbox{semi-definite} programming formulation that can be solved efficiently. Simulation results show that neglecting the CSI in the system design leads to frequent violations of the target information rate.

The \mbox{multi-user} system case is investigated in~\cite{Fouladgar2012}, where a setup with two users and a receiver is considered.
The first considerations are made on a scenario comprising a standard multiple access channel under the constraint that the energy received by the decoder is large enough. It is demonstrated that, as the required energy at the decoder increases, \mbox{time-sharing} is necessary to achieve optimal performance. This indicates the need for additional coordination between the two users.
In a second scenario, a \mbox{multi-hop} channel is considered, where the relay is assumed to be capable of harvesting the energy received from the transmitter to forward packets to the receiver. It is shown that for small SNRs in the second hop, it is desirable to maximize the energy transfer to the relay, while for sufficiently large SNRs in the second hop, it is optimal to maximize the information transfer to the relay. This means that the transmitter needs to adjust its transmission strategy according to the quality of the second link, with a subsequent need for further coordination.

A scenario with relay nodes is also considered in~\cite{Chalise2012}, where the performance limits of a \mbox{two-hop} \mbox{multi-antenna} \mbox{amplify-and-forward} relay system are investigated.
The employment of wireless energy harvesting in dense networks has been studied in \cite{Mekikis2016information}. Sensors are supplied by batteries and can harvest energy from neighbor packet transmissions. Two communication scenarios are considered: i) direct, where the sensors exchange messages directly, and ii) cooperative, where randomly deployed relays assist the message exchange. Simulation results indicate that the direct communication scenario presents better communication performance in randomly deployed dense network, whereas the cooperative scenario is superior in terms of network lifetime, providing higher harvested power.
However, the wireless energy harvesting is not able to provide enough power to counterbalance the consumed energy in realistic scenarios, mainly due to the the path loss and the RF-to-DC conversion. A solution to this problem is represented by the deployment of dedicated power transmitters of power beacons (PB) as done in \cite{Mekikis2016connectivity}, where a wireless powered sensor network with battery-less devices is considered. The authors provides results about the connectivity of the sensor network considering different routing mechanism (i.e, unicast, broadcast) and fading conditions.

\subsection{Wireless Powered Communication Network}

In this scenario, an Access Point (AP) transmits energy to multiple wireless devices. These devices use the harvested energy to transmit information in the uplink channel.
Considering a transmission block of duration $T$, during a first phase of duration $\tau_0 T$ ($0<\tau_0<1$), the wireless devices harvest energy, while in the second phase, of duration $(1-\tau_0)T$, they use the harvested energy to transmit information back to the AP. This protocol is termed \textit{harvest-then-transmit}.

A typical issue of WPCNs is defined as \mbox{\textit{doubly-near-far}} problem and it is quite similar to the \mbox{near-far} problem that was discussed in Section~\ref{subsec:WET}. In this case, a device placed further away from the AP harvests less energy than a closer device, due to the higher signal attenuation experienced by the former. For the same reason, it requires a smaller amount of power to transmit data to the AP.

A solution to this problem is proposed in~\cite{ju2014user}, where the cooperation among users is exploited in a \mbox{two-user} WPCN. The AP and the users are equipped with a single antenna. The user with the best channel, both for the EH downlink and the information transmission, uses part of its allocated uplink time and harvested energy to relay information. Simulation results show that this approach leads to improvements in the throughput and in the user fairness.

In~\cite{Liu2014}, a scenario with a \mbox{multi-antenna} AP and a number of \mbox{single-antenna} users is considered. The minimum throughput among all users is maximized (\mbox{max-min} allocation problem) by a joint design of the \mbox{downlink-uplink} time allocation, the downlink energy beamforming, the uplink transmit power allocation, and the receive beamforming, while guaranteeing fairness. An optimal \mbox{two-stage} algorithm is proposed and two suboptimal designs, exploiting \mbox{\textit{zero-forcing}} based receive beamforming, are also proposed. Numerical results show that the performance of suboptimal approaches is close to the optimal one when the distance from the AP is small, while the performance gap increases as this distance gets larger. Moreover, the \mbox{max-min} throughput is shown to increase significantly with the number of active antennas at the AP.

The same scenario, with a \mbox{multi-antenna} AP and multiple \mbox{single-antenna} users, is considered in~\cite{Yang2015}. In this paper, the transmission time frame includes a slot for the channel estimation in the uplink. The users at first consume a fraction of the harvested energy to send pilots in the uplink. Then, the AP estimates the uplink channels and obtains the downlink channel gains exploiting channel reciprocity. Hence, the scheme follows the classic steps of the \mbox{harvest-then-transmit} protocol we described above. Even though a perfect CSI at the transmitter is not available, a more accurate CSI is shown to contribute to a higher energy transfer efficiency and to lead to a higher uplink information rate.

\subsection{State-of-the-art of RF and microwave energy conversion efficiencies}
This subsection discusses the efficiencies of \mbox{state-of-the-art} energy harvesting devices, according to~\cite{Valenta2014}. The contributions on this research field have been made essentially by two communities, focusing on \mbox{space-based} solar power harvesting and \mbox{Radio-Frequency IDentification (RFID)} systems. The first community deals with energy conversion at long distances and high powers, while the second with \mbox{ultra-low} power applications. 

\begin{figure}
\input{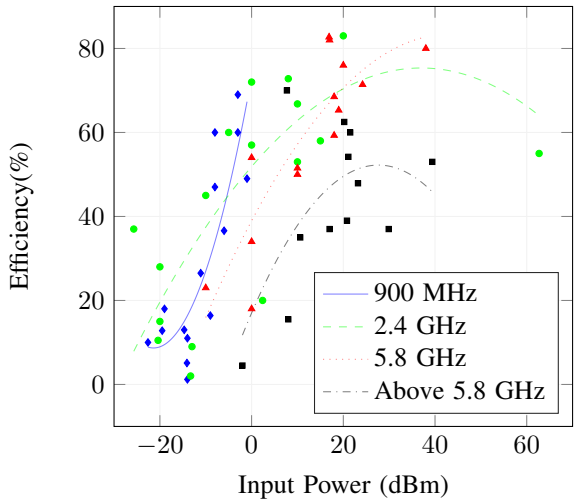}
\caption{State-of-the-art RF and microwave conversion efficiencies, from \cite{Valenta2014}. Several topologies are used under a variety of load conditions and technologies.}
\label{fig:eh_eff}
\end{figure}

In Fig.~\ref{fig:eh_eff}, the state-of-the-art efficiencies from \cite{Valenta2014} are shown. Specifically, the input power is plotted versus the energy receiver efficiency for different energy transmission frequencies, in the $\SI{900}{\mega\hertz}$, $\SI{2.4}{\giga\hertz}$ and $\SI{5.8}{\giga\hertz}$ bands. Efficiencies of applications working in bands above $\SI{5.8}{\giga\hertz}$ are also shown.

Observing the curve for the $\SI{900}{\mega\hertz}$ band, we see that these systems are typically designed to work with low input levels. This is motivated by the fact that the research on Ultra High Frequency (UHF) energy harvesters have been engineered for RFID applications. Since RFID applications are designed to work in multipath environments, the available energy levels at the receiver are low.

Other studies consider wireless power transfer applications operating at microwave frequencies. Working in these bands permits the use of smaller antennas, thus reducing the required antenna aperture and making antenna beam steering easier. In particular, the availability of the unlicensed $\SI{5.8}{\giga\hertz}$ band has led researchers to focus on it. From Fig.~\ref{fig:eh_eff}, we see that most of the microwave frequency applications discussed in the literature, operate at high power levels. This is motivated by the fact that these works consider \mbox{space-based} solar power or pure WPT applications, which typically deal with high powers.

From this graph we see that as the transmitting power increases, the efficiency of energy harvesting devices increases too, whereas it decreases with an increasing frequency. This last fact can be motivated by the higher circuitry parasitic losses encountered at microwave frequencies.

\subsection{Main outcomes}
The main findings of this section are described as follow:
\begin{enumerate}
\item The wireless transmission of energy has been studied in the literature considering different architectures, namely WET, SWIPT and WPCN.
\item The energy transmission efficiency depends on the distance between transmitter and receiver. Therefore, far users receive less energy and, in the case of WPCN, they are those that need it more to communicate. Cooperation schemes are a good solution to solve this problem.
\item Rate-Energy regions are used to find the optimal trade-off between energy transmission and information rate in SWIPT.
\item Different design for simultaneous transmission of information and power have been studied: time switching, power splitting, integrated receiver and antenna switching.
\end{enumerate}

Future information and energy networks are likely to operate on overlapping portions of the spectrum, due its scarcity. For this reason, there is a need to manage the interference that will be dominated by the transmission of energy. It is also necessary to investigate scenarios with multiple users and, specifically, new ways of mitigating the interference, such as opportunistic WET with spectrum sensing and energy/information schedulers. In particular, when considering energy transmission, harmful interference can be turned into useful harvested energy. Hence, the problem of mitigating interference while facilitating energy transfer, must be addressed.

The described literature analyses static scenarios, but nodes can also be mobile. In this case, the transmission of energy and information becomes \mbox{time-variant}, thus requiring dynamic and adaptive resource allocation policies. Further investigation is necessary to characterize the trade-off between transmit power and distance from the receiver in mobile settings.

When considering the wireless transmission of energy, the intensity of microwaves can become a problem in some areas, especially when using massive MIMO and beamforming technologies. In particular, the power radiated by wireless devices must always satisfy the Equivalent Isotropically Radiated Power (EIRP) limitations dictated by existing regulations. To solve this problem, systems based on the concept of distributed antennas can be exploited. In this way, we have an omnidirectional and weak radiation for each antenna. The combined effect of this radiation is destructive everywhere, except for the desired location, where it is constructive. This solution should be further investigated taking into account the increased power consumption due to the use of multiple antennas. In particular, the trade-off between the energy harvesting efficiency and the power consumption should be analyzed.

Finally, we underline that current studies are mainly theoretical and the achievable throughput performance for practical wireless information and energy transmission systems shall be assessed. These studies should test the use of new technologies like mmWave, massive MIMO and distributed antenna arrays.

As we already discussed, the sensitivity of the receivers is a fundamental aspect to consider in the analysis of SWIPT schemes. Actually, the low sensitivity of energy receivers represents a problem, leading to situations where a device can only decode information without harvesting energy, with the consequent degradation of the SWIPT performance. For this reason, it is necessary to improve the energy receiver circuits in terms of hardware and design.

Overall, considering (i) the energy consumption sources from Section III, (ii) the energy efficiency of WPT receivers, (iii) the limited transmission powers due to regulations and especially to (iv) wireless channel losses, WPT is not deemed an effective technology to provide energy to mobile devices, as also discussed in~\cite{Angel-WoWMoM-2017}. Transfer efficiencies are in fact very small (often smaller than $10^{-4}$) even when beamforming is exploited.

%%%%%%%%%%%%%%%%%%%%%%%%%%%%%%%%%%%%%%%%%%%%%%%%%%%%
%  ENERGY COOPERATION
%%%%%%%%%%%%%%%%%%%%%%%%%%%%%%%%%%%%%%%%%%%%%%%%%%%%

\section{Energy cooperation}
\label{sec:EnCoop}
 
We now consider a scenario where the BSs are supplied by energy harvesters and storage devices (rechargeable batteries) and may be disconnected from the power grid (\mbox{\it off-grid}). There, cooperation strategies can be conceived to make them \emph{quasi} \mbox{self-sustainable}, i.e., to operate mostly relying on the harvested (and stored) energy. 

In this context, geographical diversity shall be exploited to mitigate the \mbox{well-known} temporal and spatial variability in the energy harvesting process, especially when using renewable sources such as the wind. This aspect is partially investigated in~\cite{Chia2014}, where a network made of two BSs equipped with energy harvesters and some limited energy storage capability is considered. The authors propose an offline linear programming algorithm, which limits the power drained from the power grid when the energy profiles are deterministic. Furthermore, an online algorithm is put forward for a more realistic scenario where they are stochastic and not known a priori. As expected, the best results are achieved when the harvested energy profiles at the two BSs are sufficiently uncorrelated. In fact, if the amount of energy harvested is highly correlated, we have a problem when the energy inflow is little, as this concurrently occurs at both BSs. When the correlation is low, it is instead very likely that one BS will experience an abundant energy inflow when the other one is in a low energy state. The former BS could then transfer some of its energy to the latter. The performance gap between the two algorithms in~\cite{Chia2014} is small, reaching the minimum value for \mbox{anti-correlated} energy profiles. We observe that a low correlation in the energy profiles can be more easily reached by using different renewable types, for example solar and wind, where the latter may be very useful to mitigate the shortage of energy from solar panels during the night. 

\begin{figure}
 \centering
\subfigure[Energy sharing through a microgrid]{\includegraphics[scale = 0.8]{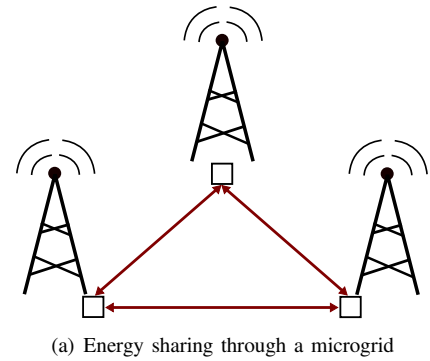}\label{fig:coop_microgrid}}
\subfigure[Energy sharing through a microgrid with an aggregator]{\includegraphics[scale=0.8]{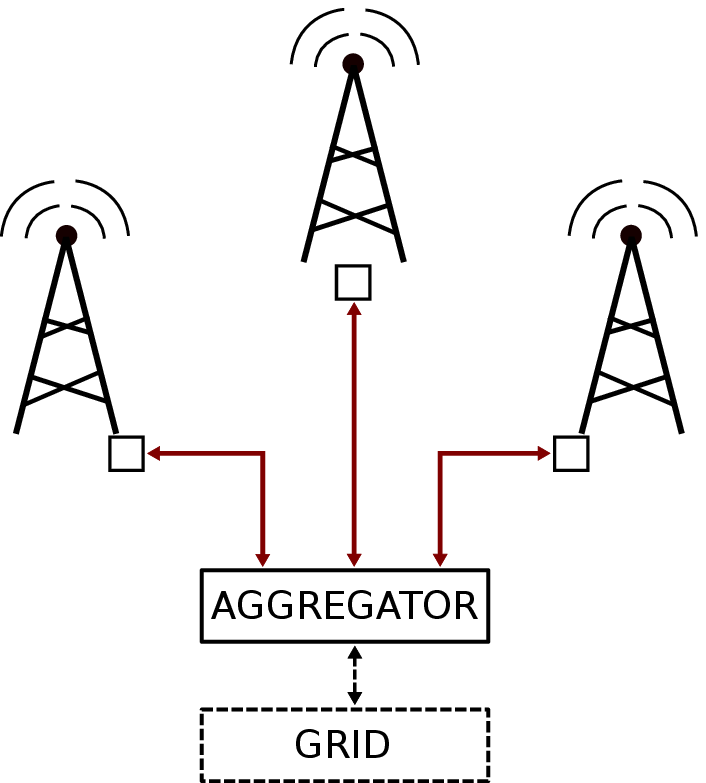}\label{fig:coop_aggregator}}
\caption{Energy sharing scenarios}  
\label{fig:coop_energysharing}
 \end{figure}

\begin{figure}
 \centering
\includegraphics[scale=0.6]{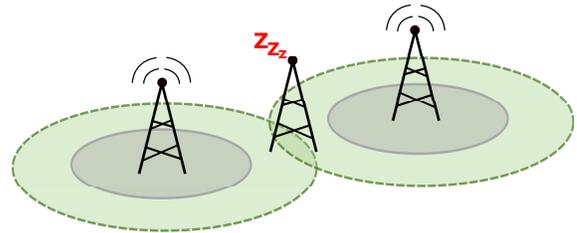}\label{fig:coop_comm}
\caption{Communication cooperation scenario}  
\label{fig:coop_communication}
 \end{figure}

In the following, two cooperation types are considered:

\begin{itemize}
\item[1)] \textbf{Energy sharing:} in this case, BSs are interconnected with electric wires, forming a sort of \mbox{microgrid} that provides mechanisms to exchange the harvested energy among the BSs. In Fig. \ref{fig:coop_energysharing} two deployment scenarios are depicted: direct connections among BSs (Fig. \ref{fig:coop_energysharing}a) and BSs connected through an aggregator (Fig. \ref{fig:coop_energysharing}b).
\item[2)] \textbf{Communication cooperation:} BSs are not interconnected via electric cables and their cooperation involves mechanisms to support the radio communication such as power control, bandwidth control, sleep modes and traffic offloading. 
In this case, \mbox{high-capacity} mmWave backhaul connections~\cite{Hur2013} can be exploited to facilitate the deployment of \mbox{\emph{drop-and-play}} devices, such as small cells. The scenario is depicted in Fig. \ref{fig:coop_communication}.
\end{itemize}

\subsection{Energy sharing}

Energy sharing among BSs  is investigated in~\cite{Gurakan2013} through the analysis of several basic multiuser network structures, namely, (i) an additive Gaussian \mbox{two-hop} relay channel with \mbox{one-way} energy transfer from the source to the relay node, (ii) a Gaussian \mbox{two-way} channel with \mbox{one-way} energy transfer and (iii) a \mbox{two-user} Gaussian multiple access channel with \mbox{one-way} energy transfer. A \mbox{two-dimensional} and directional \mbox{water-filling} algorithm is devised to control the harvested energy flows in both time and space (among users), with the objective of maximizing the system throughput for all the considered network configurations. The allocation algorithm is offline, relies on a priori information, i.e., the amount of energy harvested by sources and relays, and assumes unlimited data and energy buffers. However, these assumptions are  unrealistic.

A very interesting energy sharing framework is presented in~\cite{Gelenbe2016}, where the concept of the Energy Packet Network (EPN) is introduced. In an EPN, discrete units of energy, termed \emph{energy packets}, can be exchanged among network elements or acquired from the environment through harvesting hardware. Accordingly, the harvested energy can be modeled as a packet arrival process, the energy storage as a packet queue and the energy consumption process as a queue of loads, i.e., one or more servers. These three components of the EPN are interconnected thanks to power switches. 
Electronic systems of this type, named \emph{power packet systems}, have been recently experimented with. In some approaches~\cite{kreiner2014packetized} the packet takes the form of a pulse of current with fixed voltage and duration. Each energy packet is equipped with an encoded header, containing the information about the destination identity (i.e., its address), which is used to route the energy packet through the EPN.

The cost of deploying the \mbox{micro-grid} infrastructure that would be required by an EPN can be high. In~\cite{Gurakan2013, Zheng2014}, the use of wireless energy transfer is considered as a means to avoid the installation cost of electric cables. However, such technology has a low energy transfer efficiency nowadays, see~\cite{Valenta2014, Angel-WoWMoM-2017}.

A solution to reduce the costs of deploying electrical connections between BSs, is presented in \cite{Xu2015cost}, where a new entity named \textit{aggregator} is introduced, as shown in Fig. \ref{fig:coop_aggregator}.
The aggregator is in charge of mediating between the grid operator and a group of BSs to redistribute the energy flows. In \cite{Xu}, the authors propose an algorithm that tries to jointly optimize the transmit power allocations and the transferred energy, so as to maximize the \mbox{sum-rate} throughput for all the users. This joint communication and energy cooperation problem is proven to be convex. 
Numerical simulation shows that this approach achieves better performance than {\it no cooperation} or {\it cooperation through communication} in terms of average \mbox{sum-rate}.

Infrastructure sharing may be exploited to reduce power consumption by fairly distributing the harvested energy by the MNOs \cite{Oikonomakou2017fairness}.
The problem to capture the energy interactions among MNOs is stated as a bankruptcy game. The authors focus on the fairness among operators to further motivate cooperation. The results show that all cooperative MNOs could be provided with 6 - 7 hours of operation during non-solar hours, regardless the traffic demand. Furthermore, MNOs buy grid energy at similar percentages when no green energy is available.

\subsection{Communication cooperation}

The \mbox{micro-grid} deployment cost (i.e., the EPN installation cost) is one of the main aspects that motivate the introduction of this second cooperation mode. In this case, each BS has an energy harvester and may have a storage unit ({\it battery}), but it is not connected with the other BSs via electric cables and, in turn, cannot directly exchange energy with them, as shown in Fig. \ref{fig:coop_communication}. This approach eliminates the CAPEX related to the deployment of the \mbox{micro-grid} infrastructure (e.g., wires, converters and controllers). However, it may require harvesters and storage units with higher capacity, to achieve a certain QoS. Ongoing research aims at finding the optimal size of harvesting devices and batteries to sustain the traffic demand through the available energy budget. In particular, methods that allow the BSs to cooperatively optimize the network energy usage are proposed.

In~\cite{Zordan2015}, the fraction of time during which a BS cannot satisfy the traffic demand, due to energy scarcity is defined as \textit{outage}. The authors compute the size of harvesters and batteries as a function of the outage probability. A photovoltaic panel is considered as the harvester and the \mbox{size-outage} region is obtained for different geographical locations. The authors conclude that full network \mbox{self-sustainability} may be feasible in locations with high solar irradiation, considering the cost and dimension of the energy harvesting hardware (panels and batteries).
In~\cite{Dhillon2014}, the authors define a system model of a \mbox{$K$-tier} heterogeneous cellular network, where BSs independently switch off when their energy reserve is insufficient. The authors determine the availability region, i.e., the uncertainty in BS availability due to the finite battery capacity and to the inherent randomness in the energy harvesting process. This provides a fundamental characterization of the conditions under which standalone BSs provide the same performance as BSs relying on traditional energy sources. 
The introduction of sleeping capabilities in some BSs in order to reduce the size of their harvesting and storage devices is explored in~\cite{Marsan2013}. In this paper, sleep modes are enabled for $50$\% of the BSs, when the traffic is below $50$\% of its peak. Although simple, this scheme allows reductions in the power consumption from $10$\% to $40$\%, depending on the sleep policy, and to reduction in the size of batteries and photovoltaic panels. However, the impact of sleep modes on the user QoS is not assessed.
In~\cite{Lee2016}, an optimization problem that seeks to minimize delay and power consumption by turning off small BSs is investigated. The proposed algorithm is online and is based on the so called {\it ski rental framework}. Each agent operates autonomously at each small cell and without having any a priori information about future energy arrivals.
The algorithm is compared against a greedy scheme that uses sleep modes when the battery level is below a fixed threshold. It is shown that the proposed solution outperforms the greedy approach in terms of power consumption and network cost. The performance is evaluated assuming that energy arrivals are Poisson. This assumption is however unrealistic in most energy harvesting scenarios, as demonstrated in~\cite{Miozzo2014}, where a stochastic Markov process has been derived for solar energy harvesting systems. 

In \cite{Piovesan2017optimal}, a two-tier urban cellular network is considered, where macro BSs are powered by the power grid and energy harvesting small cells are deployed for capacity extension. The authors propose a centralized optimal direct load control of the small cells based on dynamic programming. The optimization problem is represented using Graph Theory and the problem is stated as a Shortest Path search.
The same scenario is considered in \cite{Miozzo}, where the authors propose an algorithm based on a \mbox{multi-agent} \mbox{reinforcement learning} that controls the energy spent according to the energy harvesting inflow and the traffic demand. Each node independently decides as to whether entering a sleep mode or serving the users within coverage. This algorithm is also shown to outperform a greedy scheme.

\subsection{Main outcomes}
The main findings of this section are described as follows:
\begin{enumerate}
\item Energy cooperation between BSs give better results when exploiting different types of RESs and geographical diversity.
\item Energy sharing possibilities are limited by the cost of deploying a microgrid of BSs. Some architectural solutions have been provided. In particular, the most feasible is represented by the use of an aggregator. However, EPNs represent an interesting challenge for future energy sharing deployments.  
\item Cooperation between BSs avoids the deployment of a microgrid. The dimension of energy harvesting and storage devices depends on the system outage constraints and on the deployment site.
\item BSs sleeping represents one of the most promising cooperation strategies.
\end{enumerate}

Energy cooperation is a recent and open field of research. Moreover, the definition of cooperation methods is crucial in case of energy \mbox{self-sustainability}. A key aspect is the characterization of the network load that is still not precisely captured by current analyse as already described in Section \ref{sec:ee_outcomes}.
We also underline the lack of performance assessments for the user perceived quality in the presence of energy cooperation mechanisms.

The harvesting process is usually characterized by very intensive power generation periods, interleaved with periods where the energy harvested is scarce of even absent. In the case of solar energy, for example, the generated power depends (among other things) on the season of the year. Since the system is designed for the worst case (e.g., winter months), the imbalance in the power generation across a full year may lead to an excess of energy during high power periods, which may be poorly handled. Investigations on an efficient use of the energy surplus shall be carried out to avoid this.
The impact of energy storage devices still has to be investigated. In such a case, the adoption of energy storage leads to a higher CAPEX and the trade-off between installation cost and network performance would also have to be assessed, taking into consideration the payback period.

Most of the work cited in this section solves offline optimization problems assuming a full knowledge of energy and load patterns. This is useful as a feasibility study and to obtain performance bounds, but it is still far from the design of a practical solution. In the literature, we see an increasing interest in learning and distributed approaches for the design of online algorithms. However, these control methods are not yet mapped into the proposed 5G architecture. Concepts like network softwarization and virtualization should be included in their design and their performance should be evaluated considering real traffic (user demand) and energy harvesting traces. Moreover, all the algorithms that have been published so far entail a zero delay when a BS transitions between active and sleep states.

Finally, a new research field is represented by the design of EPNs. There, energy packets would represent a flexible and convenient method to route energy when and where needed. However, the design of power switches, as well as the definition of proper energy routing protocols, are still open research directions.

%%%%%%%%%%%%%%%%%%%%%%%%%%%%%%%%%%%%%%%%%%%%%%%%%%%%
% ENERGY TRADING 
%%%%%%%%%%%%%%%%%%%%%%%%%%%%%%%%%%%%%%%%%%%%%%%%%%%%

\section{Energy trading} 
\label{sec:EnTrad}

In this section, we discuss a scenario where the 5G network trades energy with the Smart Grid (SG). In a SG communications is provided across energy producers and consumers. Energy can be bought from the main power distribution network, but also from distributed users, if equipped with some energy harvester. Finally, these users can even sell their surplus energy, injecting it into the SG. A user may then concurrently act as an energy consumer and producer (often termed \emph{prosumer} for short). In the scenario that we envision here, a BS with energy harvesting capability can be considered a {\it prosumer} of the SG.
Next, we analyze the possible interactions that may occur between the BSs of a cellular network and the SG, by reviewing the existing literature and discussing open challenges. 

\subsection{A review of energy trading in Smart Grids}

The energy supply system consists of energy {\it retailers} and {\it consumers}. The retailers offer a \mbox{source-dependent} energy price that varies over time. Consumers choose one or more retailers to buy energy from, depending on market prices.

The SG infrastructure is dimensioned to meet the peak energy demand and to avoid blackouts. This leads to an underutilization of the resources during \mbox{off-peak} periods. Furthermore, an increase in the peak demand requires investments in the distribution network and, possibly, in the power plants. For these reasons, grid operators are pushing the consumers to reduce their demand (the SG load) during peak hours (through dynamic pricing and economic incentives) or to shift their load to \mbox{off-peak} hours. The activities that target (i) reshaping the consumer's demand profile to make it match the power supply, (ii) eliminating blackouts, and (iii) reducing the operational costs and the carbon footprint are referred to as \mbox{Demand-Side} Management (DSM) in the literature. A practical way of achieving them is through Demand Response (DR), i.e., the energy provider issues some offers (incentives, etc.) over time and the users ``respond'' to these by adapting their behavior. Some researchers and practitioners use DSM and DR interchangeably~\cite{vardakas2015survey}, although DR can be seen as a way to implement DSM policies.
		
A \mbox{real-time} pricing scheme is presented in~\cite{Qian2013} to reduce the \mbox{peak-to-average} load ratio. The system is composed of several consumers and a single retailer. Each user reacts to the prices announced by the retailer and maximizes its payoff, which is the difference between its \mbox{quality-of-usage} and the cost of the energy bought from the retailer. The retailer designs realtime prices in response to the forecast user reactions to maximize its own payoff. 

Game theory, and specifically Stackelberg games, has been widely used to find distributed solutions for dynamic pricing problems. This type of game models the behavior of two agents, one of them being the leader (having the first move advantage) and the other one being the follower, who plays a best response strategy to maximize his own utility. In energy trading scenarios, the retailer is usually the leader and sets energy prices according to market needs in an attempt to spur the participation of users, while also trying to maximize its own revenue~\cite{Bayram2014}. 
A similar approach is presented in~\cite{Wei2015}, where the authors propose a decision model for a retailer, who plays the role of an intermediary agent between a wholesale energy market and \mbox{end-consumers}. The response of the consumers with respect to the retailer price follows a \mbox{two-stage} Stackelberg game, while the market price uncertainty is modeled by a robust linear optimization model. The problem is reformulated as a mixed integer linear program and solved heuristically. 
A \mbox{non-cooperative} energy supply game is formulated in~\cite{Huang2013} to capture the competitive market within a \mbox{multiple-supplier} \mbox{micro-grid}. The authors of this paper propose an iterative algorithm to find the Nash equilibrium of the energy supply game and another one to form coalitions between \mbox{micro-grids}. Their results show that the pricing mechanism reduces the electricity imbalance inside the \mbox{micro-grid} and that the profit made by choosing to cooperate is higher than that made operating independently.

Collaborative schemes among consumers, designed to reduce the energy cost, are explored in~\cite{Mohsenian2010}. There, two optimization problems are formulated with the goal of minimizing the \mbox{peak-to-average} ratio and the system energy cost. These problems are solved in a distributed manner through a scheduling algorithm based on game theory. Moreover, to encourage users to behave in a desired way (i.e., to minimize the energy cost) the authors propose a smart pricing tariff such that the interactions among the users automatically lead to an optimal aggregate load profile.
Cooperation is also investigated in \cite{Zenginis2017cooperation} for the case of urban buildings composing a \mbox{micro-grid}. The problem of deciding the optimal capacities of the harvesting equipment as well as of determining the optimal daily power operation plan is formulated as a mixed integer linear program. The objective function is optimized based on the Nash bargain method to enable equally distributed savings among the participants. Results show that power exchanges affect the required equipment size and viceversa. Furthermore, energy exchanges enhance the system \mbox{self-sufficiency} and reduce carbon emissions.

In~\cite{Dong2012}, the \mbox{social-welfare}, defined as the difference between the total demand, the total cost experienced by all the generators and the wastage cost caused by transmission losses, is maximized through a distributed {\it demand and response} algorithm. The problem is formulated using convex optimization and solved in a distributed fashion applying the \mbox{Lagrange-Newton} method. In the computation of the optimal solution, each node (consumer or supplier) exchanges rounds of messages with its neighboring nodes. Although simulation results verify the correctness of the distributed algorithm, the computation rate and the entailed communication load are rather high.
DSM is also a viable approach to control the temporal separation between energy generation and demand. In fact, load shifting allows demand flexibility without compromising the QoS~\cite{lund2015review}. This flexibility can be achieved thanks to energy storage devices, which can be used to accumulate renewable energy and use it when needed.

\subsection{Cellular networks meet SG}
The interaction between cellular networks and the SG can be implemented in two different ways: 1) the SG is the only energy supplier and 2) the energy harvesting BSs and the SG are the energy suppliers.\\

\mypar{1) The SG is the only energy supplier:} in this scenario, several retailers operate within the SG to serve the consumers. An example of this can be found in~\cite{Zhang2014}, where a power allocation scheme, formulated as a \mbox{non-cooperative} game, is put forward to increase the network energy efficiency. Retailers offer different prices to the BSs and a \mbox{multi-agent} \mbox{Q-learning} scheme is proposed for the game to reach the optimal transmission power configuration. Along the same lines, in~\cite{Bu2013} a cognitive HetNet only powered by the SG is considered. In this paper, the authors formulate the problems of: (i) electricity price decision, (ii) \mbox{energy-efficient} power allocation and (iii) interference management, which are jointly and iteratively solved as a \mbox{three-level} Stackelberg game.

In~\cite{farooq2016stochastic}, a DSM framework for cellular networks powered by multiple energy suppliers is proposed. The system model comprises a set of cellular operators, characterized by the QoS offered to their subscribers, and powered by a common pool of energy suppliers, characterized by energy prices and pollutant emission levels. \mbox{Closed-form} expressions for the amount of energy provided by each supplier to the operators are derived using stochastic geometry, accounting for user QoS, energy cost and carbon emissions.\\

\begin{figure}[t]
	\centering
	\subfigure[One supplier]{\includegraphics[]{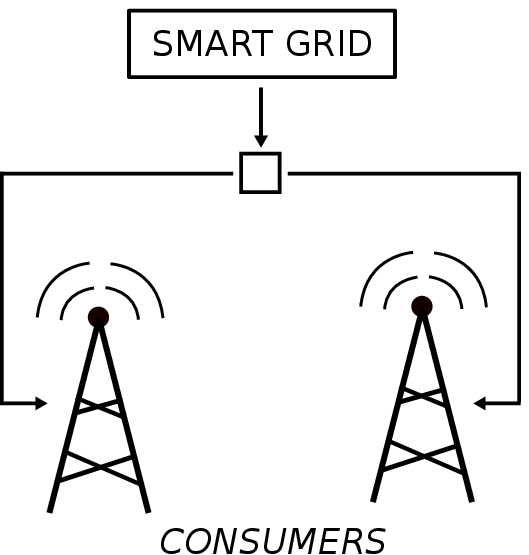}\label{fig:trade_1}}
	\subfigure[Multiple suppliers]{\includegraphics[width=\columnwidth]{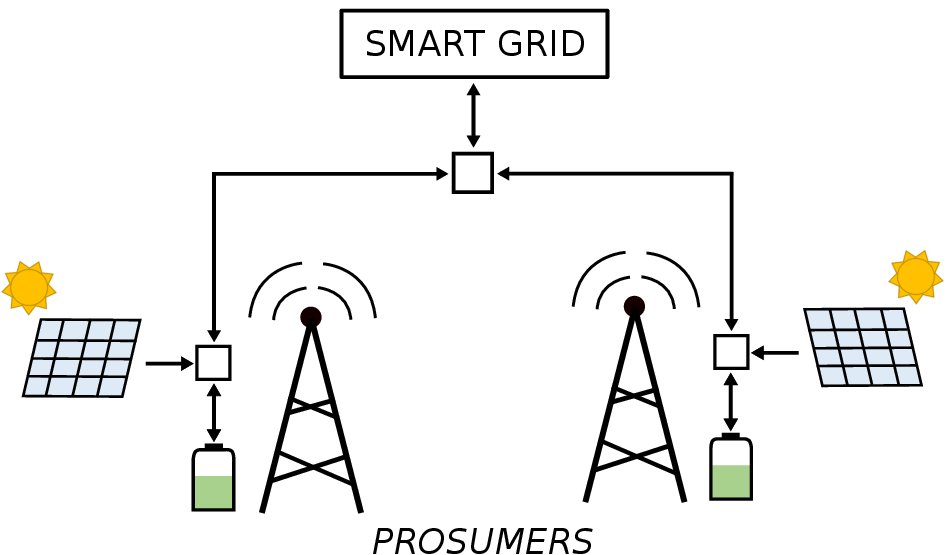}\label{fig:trade_2}}
	\caption{Diagram illustrating the two ways of interaction between mobile networks and the smart grid.}
	\label{figure:trade}
\end{figure}

\mypar{2) The suppliers are the SG and the energy harvesting BSs:} in this second scenario, BSs have energy harvesting capabilities, and act as {\it prosumers} of the SG, see Fig.~\ref{figure:trade}. Different scenarios can be envisioned. For example, in~\cite{Ghazzai2014} mobile operators are responsible for supplying power to their BSs. Each network operator has to procure energy from several SG retailers. The procurement decision is affected by two factors: the unitary price of energy and a penalty term depending on the amount of pollutant emissions from the energy source. Moreover, BSs are prosumers, i.e., they can procure energy from their own RESs, which are free of charge for the network operator. Using a \mbox{two-level} Stackelberg game, the authors of~\cite{Ghazzai2014} formulate and find the optimal solution for an optimization problem that seeks to maximize the operator profit, as well as to reduce the emission of pollutants.
We remark that, besides this centralized \mbox{decision-making} model, where the network operator decides the energy retailer for each of BSs, there are distributed scenarios where BSs are themselves responsible for carrying out the acquisition of energy in a distributed manner, choosing the most appropriate retailer according to their energy status, i.e., on their current energy income and reserve. 

An adaptive power management for wireless BSs is studied in~\cite{Niyato2012}. Here, each BS is a prosumer equipped with a solar panel and an energy storage unit, but is also plugged into the electrical grid. Due to the random nature of renewable generation, power prices and traffic load, the authors formulate a \mbox{multi-stage} stochastic optimization problem. This problem is then framed as a linear program and solved using standard tools.
Energy management strategies are presented in~\cite{Leithon2014, Leithon2014conference}. In these works, the authors also elaborate on the use of storage devices. In~\cite{Leithon2014}, simulations results show that a cost reduction can be attained through a higher battery capacity, but a greater cost reduction is possible by increasing the number of BSs. In~\cite{Leithon2014conference}, the authors get a critical battery capacity level above which no further cost reduction can be achieved. According to the authors, these results can be used as guidelines in the design of storage systems for BSs in a SG environment.
Auctioning is explored in~\cite{Reyhanian2016}, where a double auction trading algorithm is proposed to incentivize BSs with extra harvested energy to share their energy surplus with BSs with a lower energy reserve. Auction mechanisms are the key elements of many applications in wholesale and retail electric power markets. Similar to traditional auction rules, the main goal of distributed energy trading is to find the \mbox{lowest-cost} match between the supply and the demand, so as to maximize the economic efficiency~\cite{Bayram2014}. 
BS energy storage is studied in~\cite{Mendil2016}. Batteries must operate within a guard range to avoid a rapid decrease of their performance (i.e., typically between the $20\%$ and $90\%$ of their capacity). The authors propose a fuzzy \mbox{Q-Learning} small cell energy controller to simultaneously minimize the electricity bought from the SG and enhance the life span of the storage device.

Finally, a hybrid energy sharing framework is proposed in \cite{farooq2017hybrid}, where a combination of physical power lines and energy trading with other BSs using smart grid is used. Algorithms for physical power lines deployment between BSs are designed based on the renewable energy availability. An energy management framework is also formulated to optimally determine the quantities of electricity and renewable energy to be procured and exchanged among BSs, respectively. Results demonstrate considerable reduction in average energy cost thanks to the hybrid energy sharing scheme.

	\subsection{Main outcomes}
	The main findings of this section are described as follows:
	\begin{enumerate}
		\item Collaborative schemes and energy cooperation among consumers in smart grids are effective techniques to reduce energy costs, while increasing network efficiency. Specifically, game theory and auctioning schemes have been widely investigated within DSM strategies, providing valuable outcomes for energy trading. Stackelberg games are the most popular approach.
		\item The use of BSs with energy harvesting capabilities opens new scenarios in the smart grid market, where green network operators could trade their harvested energy with the SG.
		\item Some initial papers dealing with the interaction between SGs and green network operators (managing BS with energy harvesting capabilities) have recently appeared. Initial results, for selected network scenarios, look promising. In particular, energy resources can be optimally allocated (and traded between BSs and the SG) to obtain monetary cost reductions and a higher energy efficiency for the BS network.
	\end{enumerate}

A few open issues are now identified. 
The basic scenario studied in the reviewed literature involves a single retailer (the SG), which offers hourly energy prices to the final consumers, i.e., the BSs. The energy price depends on the cost of production and on the expected demand. In this scenario, \mbox{decision-making} solutions shall be addressed to find the best \mbox{energy-purchasing} policies for the BSs taking into account: (i) current and forecast renewable energy income, (ii) current and forecast traffic load and (iii) the future evolution of the energy prices. In addition, the presence of energy storage devices makes the problem more involved, allowing the storage of energy for later use, when the market conditions are unfavorable. There is a vast literature on dynamic pricing and price forecast, but this is mostly limited to the smart grid domain, whereas the integration of prices, energy and load forecast for the control of BSs, when these act as {\it prosumers} within a SG, is still unexplored.

Further, existing papers study network scenarios where the BSs can harvest energy, use it locally (to serve their own mobile users) or purchase it from the SG. Few studies additionally consider BSs as possible energy sources, and allow them to sell energy to the SG retailer. However, more complex scenarios are possible, where BSs interact and are endowed with the capability of exchanging energy among themselves (using their local energy storage). According to this new paradigm, BSs can sell(buy) energy to(from) other BSs in the mobile network, besides using it locally or selling it to the main SG retailer. This amounts to green mobile networks, where BSs can \mbox{self-organize} and cooperate toward the overall reduction of the energy that the mobile network drains from the SG, reducing the carbon footprint of ICT.

Finally, stochastic optimization and adaptive control tools, involving, e.g., model predictive control, shall be considered to handle the integration of energy harvesting capabilities in mobile networks where BSs can be considered as prosumers of the SG. Better load models (accurately tracking the spatio-temporal traits of mobile traffic) are needed, along with lightweight and flexible tools for pattern analysis and prediction, to be integrated into foresighted optimization techniques. Within these settings,  \mbox{micro-economic} models should also be investigated when consumers (BSs) aim at maximizing their utility (e.g., combining energy, monetary cost and served traffic), subject to their monetary budget constraints; while SG retailers aim at maximizing their profit.

\section{Conclusions}
\label{sec:conclusions}

In this survey, we have elaborated on the use of energy harvesting hardware as a means to decrease the environmental footprint of 5G technology. To take full advantage of the harvested (renewable) energy, while still meeting the quality of service required by dense 5G deployments, suitable management techniques have been reviewed, highlighting the open issues that are still to be solved to provide \mbox{eco-friendly} and \mbox{cost-effective} mobile networks. While several techniques have recently been proposed to tackle capacity, coverage and efficiency problems, including: C-RAN, software defined networking and fog computing, these are deemed insufficient and do not generally consider network elements possessing renewable energy harvesting capabilities. From the analysis that we have carried out in this survey, we have identified several open issues that range from the need for accurate energy, transmission and consumption models, to the lack of accurate data traffic profiles (from real mobility traces), to the use of power transfer, energy cooperation and energy trading techniques. Specifically, current wireless transfer techniques are deemed inadequate to provide energy balancing across network elements, mainly due to their very small transfer efficiencies. Energy cooperation techniques look very promising and should be better addressed, including energy harvesting and traffic dynamics. In this respect, energy packet networks are envisaged to be an interesting solution to be further explored for energy transfer among network nodes.  
Energy trading is finally assessed, looking at renewable mobile networks as part of electricity grids and at their network elements as possible prosumers. In a near future, in fact, energy may also be sold to electricity providers or utilities. Dedicated market rules shall be identified to this end.

\section*{Acknowledgment}
This work has received funding from the European Union's Horizon 2020 research and innovation programme under the Marie \mbox{Sklodowska-Curie} grant agreement No 675891 \mbox{(SCAVENGE)}.

\bibliographystyle{elsarticle-num}
\bibliography{survey}

\end{document}